\documentclass[preprint, floatfix
]{revtex4-2}
\usepackage[utf8]{inputenc}
\usepackage{standalone}
\usepackage{graphicx}
\usepackage{dcolumn}
\usepackage{multirow}
\usepackage{diagbox}
\usepackage{bm}
\usepackage{braket}
\usepackage{dsfont}
\usepackage{color} 

\usepackage[normalem]{ulem}
\usepackage{soul}
\usepackage{svg}
\usepackage{mathtools} 

\usepackage[breaklinks, hidelinks]{hyperref}
\usepackage{xcolor}
\hypersetup{
    colorlinks,
    linkcolor={red!80!black},
    citecolor={red!80!black},
    urlcolor={red!80!black}
}

\usepackage[caption=false]{subfig}

\graphicspath{{figures/}}
\def\doteq{\,\overset{\boldsymbol{.}}{=}\,}

\begin{document}

\title{Optical Response Beyond Magnetic Symmetries}

%

\date{\today}
\author{Javier Sivianes}
\affiliation{Centro de F\'{i}sica de Materiales (CSIC-UPV/EHU), 20018, Donostia-San Sebasti\'{a}n, Spain}

\author{Enrique Boquete-Someso}

\affiliation{Centro de F\'{i}sica de Materiales (CSIC-UPV/EHU), 20018, Donostia-San Sebasti\'{a}n, Spain}

%
\author{Daniel Hernangómez-Pérez}
\affiliation{CIC nanoGUNE BRTA, Tolosa Hiribidea 76, 20018 San Sebasti\'an, Spain}
\author{Julen Iba\~{n}ez-Azpiroz}%
\affiliation{Centro de F\'{i}sica de Materiales (CSIC-UPV/EHU), 20018, Donostia-San Sebasti\'{a}n, Spain}
\affiliation{IKERBASQUE, Basque Foundation for Science, 48009 Bilbao, Spain}
\affiliation{Donostia International Physics Center (DIPC), 20018
Donostia-San Sebasti\'{a}n, Spain}

\begin{abstract}
The optical response of magnetic materials is conventionally classified through magnetic space groups (MSGs), where spin and lattice are locked by the relativistic spin-orbit interaction. However, most optical observables are governed primarily by nonrelativistic physics, and thus a purely MSG-based description can overlook 
important insights.
%
%
Here we systematically show that spin-space groups (SSGs), 
which operate at the nonrelativistic level, 
provide a broader and more predictive framework for analyzing 
a variety of optical responses of magnets. 
Focusing on linear optical absorption, we derive the transformation rules imposed by SSGs and show that they generate effective real-space point groups, which can enforce relations among charge response coefficients that are  absent from conventional MSG analysis.
We illustrate the basic principle in a Lieb-lattice altermagnet model with tunable spin-orbit coupling, where SSG predictions on the linear dichroism remain remarkably accurate even when relativistic band splittings become sizable.
%
%
%
We further establish the predictive power of  this framework  through first-principles calculations on two altermagnetic candidates: the actinide UCr$_2$Si$_2$C, where the optical absorption remains nearly isotropic despite its strong spin-orbit coupling and
the pronounced anisotropy apparent from magnetic symmetries, and the 
transition-metal fluoride 
RbMnF$_4$, where  birefringence is confined to a single plane
by symmetries emerging exclusively from SSGs. 
%
Finally, we extend the concept to the spin Hall response of the coplanar noncollinear antiferromagnet ScMnO$_3$, 
where SSGs explain the hierarchy  of calculated spin Hall coefficients, demonstrating their direct relevance to spintronics as well.
%
Overall, our results establish SSG analysis as a powerful framework for understanding and predicting charge and spin optical responses in magnetic materials, providing valuable guidance for both theoretical studies and the interpretation of experimental optical spectra.

\end{abstract}

\maketitle

\section{Introduction}

Symmetry principles play a central organizing role in condensed matter physics, providing a unifying framework for classifying phases of matter and understanding their emergent properties.
Magnetic order is a particularly clear example of this paradigm, since the classification of magnetic materials is fundamentally rooted in their symmetry structure.
Magnets have long been classified~\cite{10.1093/oso/9780198505921.001.0001}, at the basic level, as either ferromagnets, in which magnetic moments align parallel in real space, or antiferromagnets, characterized by compensated antiparallel alignment. 
The traditional approach to this categorization is based on the symmetry description of magnetic crystal structures, which considers transformations acting jointly in real space and in the space of magnetic moment vectors, forming so-called magnetic space groups (MSGs)~\cite{Litvin2008-ka}.
%
These coupled operations, however, are intrinsically relativistic in origin, since the leading microscopic interaction linking spin and real space is the spin-orbit coupling (SOC).
Therefore, while MSGs are essential for capturing relativistic effects, they are less well suited for classifying nonrelativistic behavior,  
which is typically the primary driver of magnetic phenomena.


To address this limitation, the use of \emph{spin space groups} (SSGs) has been advocated in recent work~\cite{PhysRevX.12.031042,PhysRevX.12.040501}.
SSGs extend the concept of conventional MSGs by lifting the constraint that ties spin rotations to real-space symmetry operations. As a result, their symmetry structure is substantially more general, as  independent rotations may act concurrently in spin  and  real space~\cite{PhysRevX.14.031038,PhysRevX.14.031037,Etxebarria:cam5007}. 
Remarkably, this symmetry description offers a natural explanation for  unconventional antiferromagnets exhibiting pronounced yet fully compensated spin splittings~\cite{Hayami_2019,PhysRevB.102.014422,Ma2021}, now known as altermagnets~\cite{PhysRevX.12.031042,PhysRevX.12.040501}. 
In recent years, SSGs have also enabled the classification and prediction of novel topological states beyond the scope of conventional MSG analysis~\cite{PhysRevX.14.031037}, the identification of unconventional magnetic phases such as 
surface altermagnets~\cite{mkk5-b53m,hw4l-jknk,leeb2026topologicallyprotectedsurfacealtermagnetism,lange2026emergentaltermagnetismsurfacesantiferromagnets}, 
odd-parity magnets~\cite{hellenes2024pwavemagnets}
and their nonlinear optical response~\cite{PhysRevLett.134.196907}, 
or developments on Landau theories of altermagnetism~\cite{PhysRevLett.132.176702,q44z-ynbr}.

Nonrelativistic SSGs are a prototypical example of approximate, or so-called hidden, symmetries. Their implications are not strictly exact, since relativistic effects, however weak, are inevitably present in real materials. Their importance lies in the fact that they nevertheless can provide robust predictions for physical phenomena that are not dominated by relativistic interactions. 
The main focus of this paper, the optical response, 
is among the most fundamental properties of matter and belongs to this category.
Despite the growing interest in SSGs, their implications in the optical properties of crystals are still largely unexplored,
with few recent exceptions, for instance in the context of excitons~\cite{zn7r-k1xd,cqn4-lljy,tdrm-twnt}. 
Establishing how spin groups constrain the optical tensors therefore represents a conceptual extension of symmetry-based response theory and a practical route toward identifying experimental signatures of these hidden symmetries.
%

In this work, we establish the predictive power of SSGs for the optical response of magnetic materials through a comprehensive classification of the resulting symmetry constraints and their systematic first-principles validation. We focus primarily on altermagnets, whose intrinsic spin symmetry structure provides an ideal platform for uncovering symmetry relations on optical response tensors that remain invisible within the conventional magnetic-group framework.

The paper is organized as follows. In Sec.~\ref{sec:symmetries}, we introduce the symmetry framework and derive the transformation rules of the response tensors.
In Sec.~\ref{sec:lieb}, we analyze the main effects in an altermagnetic Lieb lattice, which serves as a minimal model for investigating the influence of spin-orbit coupling on the optical predictions derived from spin-group symmetries. Finally, in Sec.~\ref{sec:dft}, we present density functional theory (DFT) calculations for representative magnetic materials exhibiting a range of optical phenomena predicted by spin groups. Technical details of the calculations are provided in the Appendix.






%



\section{Magnetic versus spin groups in optical coefficients}
\label{sec:symmetries}

\subsection{Background and transformation properties of response tensors}
The symmetry of nonmagnetic crystals can be described by space groups, whose elements are denoted as $g_M=\{R|\mathbf{t}\}$, where $R$ represents a proper or improper rotation and $\mathbf{t}$ is a fractional translation.
When accounting for magnetic order, the symmetry description must be extended to include time-reversal symmetry, represented by $\mathcal{T}$, forming MSGs (also known as Shubnikov groups~\cite{Lifshitz_2024, chakraborty2023encyclopedia}). 
In these groups, operations may combine spatial transformations with $\mathcal{T}$. Importantly, the transformation $R$ must preserve both the atomic positions and the orientation of the magnetic moment, which generally leads to a reduction in the available symmetries when compared to the nonmagnetic case.
The role of SOC becomes crucial in determining the ``true'' magnetic symmetries, as it couples orbital and spin degrees of freedom. Consequently, any symmetry operation must simultaneously and consistently transform both spatial coordinates and spin orientations.

In the nonrelativistic limit of vanishing SOC, the symmetries of the Hamiltonian are no  longer  restricted to the magnetic group. This is because now independent rotations in spin and real space are allowed, leading to the emergence of SSGs.
These symmetries were first introduced by Brinkman and Elliott \cite{doi:10.1098/rspa.1966.0211, 10.1063/1.1708514} and later generalized by Litvin and Opechowski \cite{LITVIN1974538} and have recently gained significant attention in condensed matter physics. 
Spin group operations can be denoted as $g_S = \{U\|\{R|\mathbf{t}\}\}$, where $U$ acts on spin space and $\{R|\mathbf{t}\}$ on real space.
Their action on the electronic charge density $\rho(\mathbf{r})$ and the spin magnetization density $\mathbf{S}(\mathbf{r})$ is given by \cite{Etxebarria:cam5007}
\begin{align}
    \rho(\mathbf{r}) &\xrightarrow{g_S} \rho(\{R|\mathbf{t}\}^{-1}\mathbf{r}), \\
    \mathbf{S}(\mathbf{r}) &\xrightarrow{g_S} U \mathbf{S}(\{R|\mathbf{t}\}^{-1}\mathbf{r}).
\end{align}

The distinction between magnetic and spin symmetries becomes particularly clear when examining their effects on material response tensors.
Since the translational component of magnetic/spin symmetries does not constrain the response tensors, in the following discussion only point group symmetries will be involved.
We illustrate this distinction using the two physical quantities that form the focus of this paper.

\subparagraph{\small{Optical photoconductivity.}} The linear optical conductivity tensor, with tensor components $\sigma^{ab}(\omega)$, relates vector components of the current density, $j^{a}(\omega)$, to those of the electric field, $E^{b}(\omega)$, through the constitutive relation 
\begin{equation}\label{eq:linear-current}
j^{a}(\omega) = \sigma^{ab}(\omega)E^{b}(\omega).    
\end{equation}
As a  second-rank tensor, $\sigma^{ab}(\omega)$ can be decomposed into its symmetric and antisymmetric parts:
\begin{align}
\sigma_{S}^{ab}(\omega) = [\sigma^{ab}(\omega) + \sigma^{ba}(\omega)]/2,\\
\sigma_{A}^{ab}(\omega) = [\sigma^{ab}(\omega) - \sigma^{ba}(\omega)]/2.\label{eq:sigma_A}
\end{align}
Considering the Onsager relations \cite{Landau1984}, it is easy to show that the symmetric part $\sigma_{S}^{ab}(\omega)$ is even under time-reversal ($\mathcal{T}$-even), while the antisymmetric part $\sigma_{A}^{ab}(\omega)$ is odd ($\mathcal{T}$-odd).
The diagonal components $\sigma_{S}^{aa}(\omega)$  are primarily responsible for the absorption of linearly polarized light, capturing effects such as linear dichroism~\cite{PhysRevB.104.235403}. In contrast, the off-diagonal components of $\sigma_{S}^{ab}(\omega)$ describe birefringence, \textit{i.e.}, the dependence of a material’s refractive index on the polarization of light~\cite{angle}.
As for the antisymmetric component $\sigma_{A}^{ab}(\omega)$, it is off-diagonal by construction and governs phenomena associated with circular dichroism, such as Kerr and Faraday rotations, as well as the Hall conductivity in the DC limit~\cite{doi:10.1126/science.1248552}. 


Under a magnetic symmetry operation, the optical conductivity transforms as 
\begin{align}
    \sigma_{S}^{ab} \xrightarrow{g_M}& R_{aa'} R_{bb'} \sigma_{S}^{a'b'}, \label{eq:spin_dielectric}\\
    \sigma_{A}^{ab} \xrightarrow{g_M}& \theta R_{aa'} R_{bb'} \sigma_{\textcolor{black}{A}}^{a'b'},
\end{align}
where $\theta = \pm 1$ depending on whether time reversal is included in the symmetry operation ($\theta = +1$) or not ($\theta = -1$).
For spin symmetries, the corresponding transformation reads
\begin{align}
    \sigma_{S}^{ab} &\xrightarrow{g} R_{aa'} R_{bb'} \sigma_{S}^{a'b'}, \label{eq:transform_sigma_S} \\
    \sigma_{A}^{ab} &\xrightarrow{g} \det(U) R_{aa'} R_{bb'} \sigma_{A}^{a'b'}.
\end{align}
At first sight, the action of SSGs may appear equivalent to that of MSGs, since the optical conductivity does not explicitly depend on spin. However, an important distinction arises because the real-space rotations $R$ entering the SSGs are more general than the symmetry operations contained in the corresponding MSGs. As a consequence, purely charge-response tensors such as $\sigma_{S}^{ab}$ and $\sigma_{A}^{ab}$ transform according to an effective point group determined solely by the spatial sector of the SSG, thereby imposing stronger constraints on the response coefficients than those obtained from MSG alone.



\begin{figure*}
    \centering
    \includegraphics[width=1\linewidth]{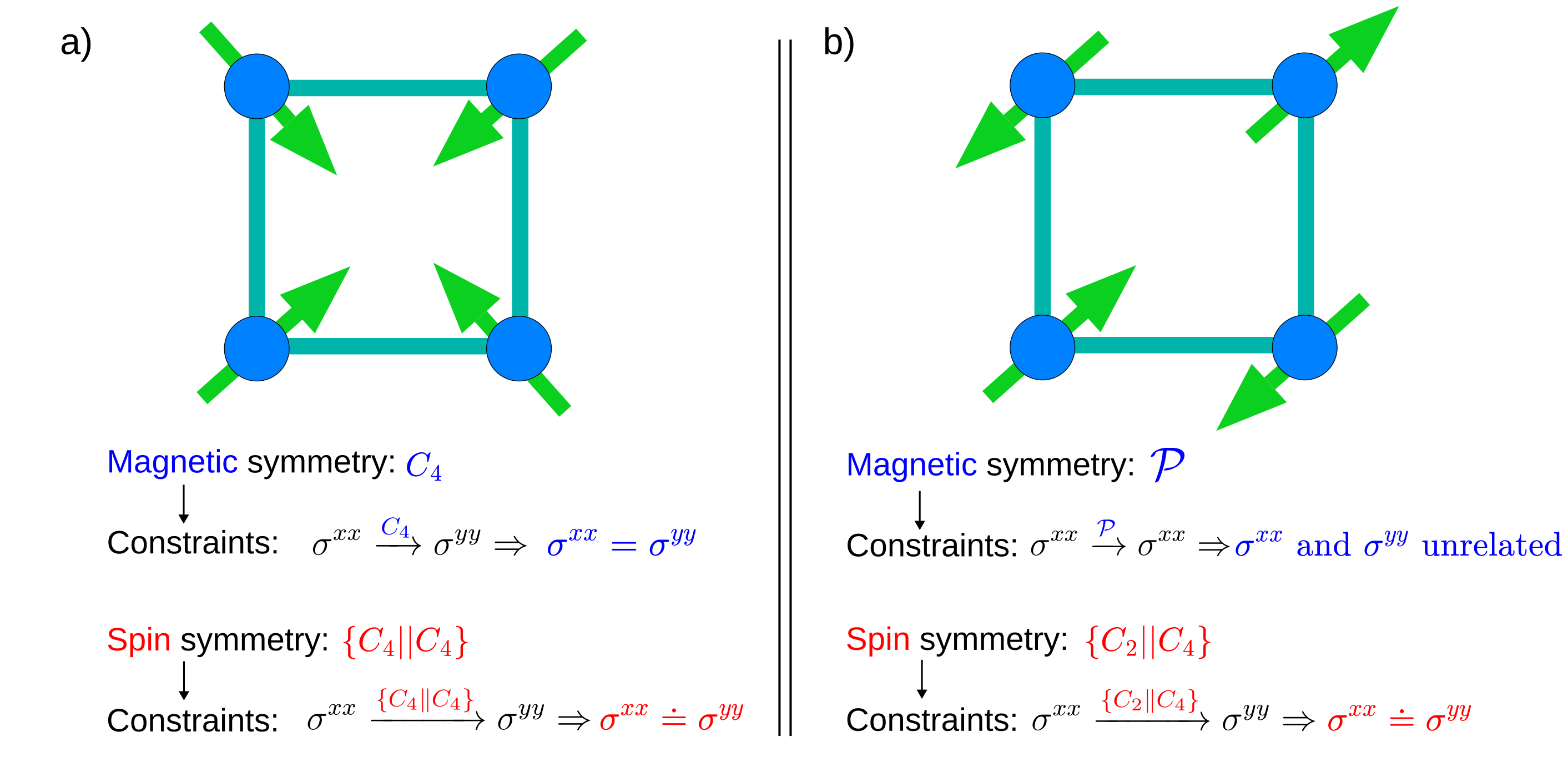}
    \caption{Two square configurations with different
    orientations of the magnetic moments. 
    In both cases, the corresponding magnetic and spin symmetries 
    as well as their constrains on the diagonal optical 
    coefficients are denoted. For the magnetic group, the structure on the 
    left imposes the exact relation $\sigma^{xx}=\sigma^{yy}$, whereas for the structure on the right
    the two coefficients are unrelated.
    For the spin group, both structures impose the ``non-relativistic'' 
    constraint $\sigma^{xx}\doteq\sigma^{yy}$.
    }
    \label{fig:spin_cartoon}
\end{figure*}

The above concept is illustrated through the two-dimensional toy models shown in Fig.~\ref{fig:spin_cartoon}, which depict two square configurations with compensated magnetic moments. In Fig.~\ref{fig:spin_cartoon}a, the magnetic configuration preserves $C_4$ symmetry in the MSG sense. 
As a consequence, the $x$ and $y$ directions remain symmetry-equivalent, enforcing the constraint $\sigma^{xx} = \sigma^{yy}$ on the diagonal optical coefficients. In contrast, in Fig.~\ref{fig:spin_cartoon}b, the magnetic moments break the $C_4$ magnetic symmetry, so that no MSG operation relates the $x$ and $y$ directions and the coefficients $\sigma^{xx}$ and $\sigma^{yy}$ become independent. Nevertheless, the system remains invariant under the spin-group operation $\{C_2||C_{4}\}$, namely a 180$^{\circ}$ spin rotation followed by
a 90$^{\circ}$ spatial rotation around the $z$ axis. Since the optical conductivity tensor is insensitive to the spin-sector component of the spin symmetry, the associated real-space operation $C_4$ still enforces the relation $\sigma^{xx} \doteq \sigma^{yy}$, where we used the symbol $\doteq$ to denote equalities implied by spin-group symmetries.



\subparagraph{\small{Spin photoconductivity.}}
The spin conductivity $\sigma^{s,ab}(\omega)$ describes the generation of a spin-polarized current component $j^{s,a}(\omega)$, with spin polarization along the axis $s=x,y,z$, in response to an applied electric field $E^{b}(\omega)$,
\begin{equation}\label{eq:spincurrent}
j^{s,a}(\omega) = \eta^{s,ab}(\omega)E^{b}(\omega).
\end{equation}
Unlike the optical conductivity,  this current is spin-resolved and the spin index now appears explicitly.
In this case, Onsager relations imply that the antisymmetric component $\eta_{A}^{s,ab}(\omega)$ is $\mathcal{T}$-even, while the symmetric component $\eta_{S}^{s,ab}(\omega)$ is $\mathcal{T}$-odd.
Under a magnetic symmetry operation, the spin conductivity tensor transforms as
\begin{align}
    \eta_A^{s,ab} &\xrightarrow{\{R|\mathbf{t}\}}  R_{ss'} R_{aa'} R_{bb'} \eta_{A}^{s',a'b'}, \\
    \eta_S^{s,ab} &\xrightarrow{\{R|\mathbf{t}\}} \theta R_{ss'} R_{aa'} R_{bb'} \eta_{S}^{s',a'b'}.
\end{align}
For spin symmetries, such transformation reads
\begin{align}
    \eta_A^{s,ab} &\xrightarrow{g} U_{ss'} R_{aa'} R_{bb'} \eta_{A}^{s',a'b'}, \label{eq:spg_shc_a} \\
    \eta_S^{s,ab} &\xrightarrow{g} \det(U) U_{ss'} R_{aa'} R_{bb'} \eta_{S}^{s',a'b'}. \label{eq:spg_shc_s}
\end{align}
Note that in the equations above, the explicit appearance of the spin rotation matrix, $U$, acting on the spin index $s$, highlights a fundamental distinction between SSGs and MSGs in the spin conductivity tensor.
Particularly, spin symmetries impose stronger constraints on  tensor components in collinear magnetic materials, as $U$encodes the rich symmetry structure of the spin configuration~\cite{Etxebarria:cam5007}. 




\section{Lieb lattice model}
\label{sec:lieb}

We begin our quantitative analysis by considering a modified 2D Lieb lattice model, which provides a simple model for describing altermagnetic spin-split states. Our main objective is to identify trends in the response as a function of SOC, which can be tuned in a controlled manner within the model. We focus exclusively on the optical response, as the 
spin Hall conductivity is heavily constrained by the collinearity of the model.


\subsection{Setup and ground state properties}

\begin{figure*}
    \centering
    \includegraphics[width=\linewidth]{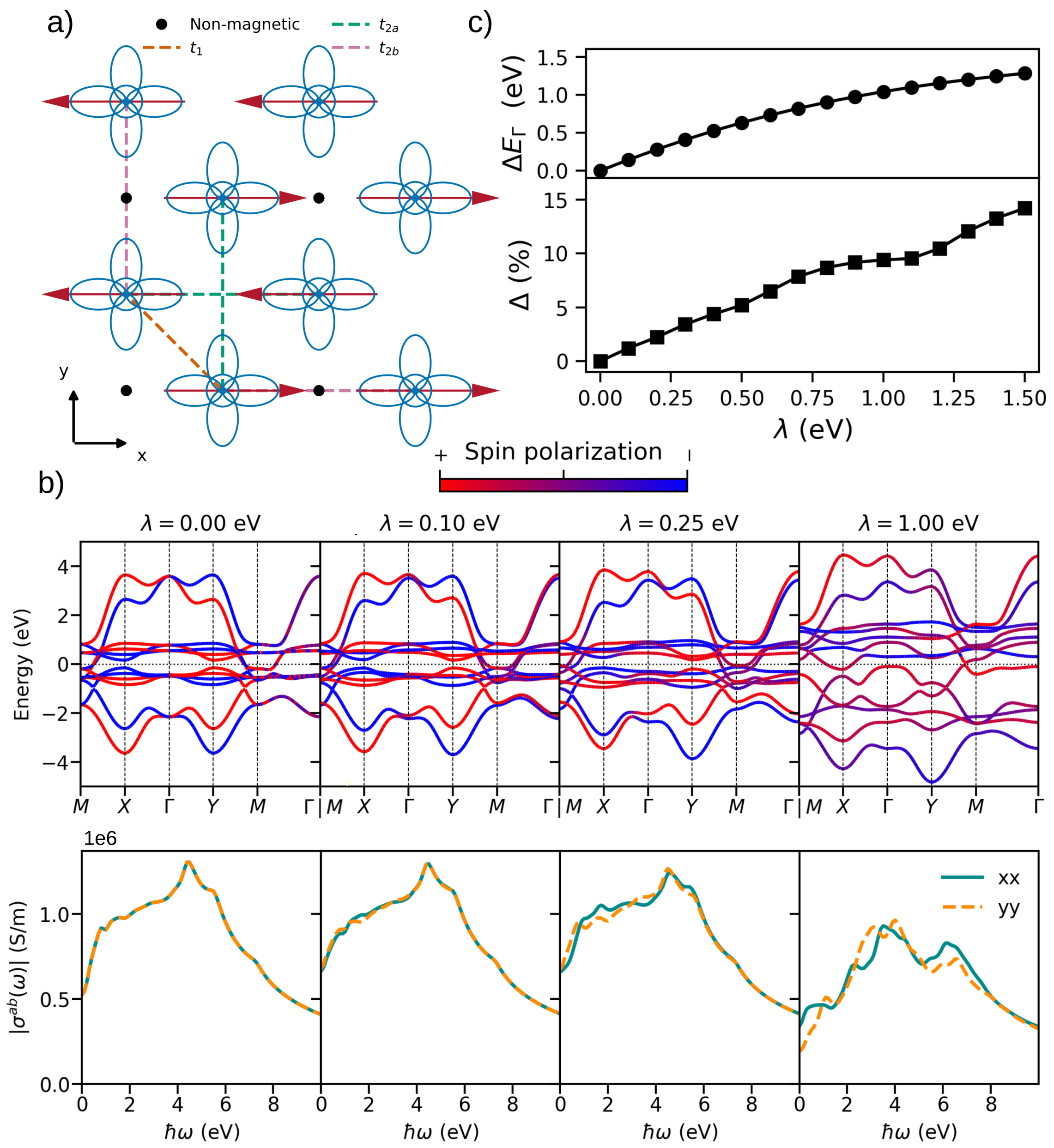}
    \caption{a) Lieb lattice with magnetic moments and tunnel couplings highlighted. 
             b) Band structure and optical conductivity for different values of the spin orbit coupling $\lambda$. 
             c) Evolution of the spin splitting $\Delta E_{\Gamma}$ and the linear dichroism metric $\Delta$ for increasing values of $\lambda$.}
    \label{fig:model}
\end{figure*}

We consider the altermagnetic Lieb lattice tight-binding model inspired by Refs.~\cite{PhysRevB.108.224421,PhysRevLett.134.096703}.
%
%
The lattice structure, depicted in Fig.~\ref{fig:model}a, consists of a square lattice with atomic sites at positions $(0,0)$, $(0.5,0)$, and $(0,0.5)$ in the unit cell (expressed in units of lattice parameter).
The site at $(0,0)$ is nonmagnetic, while the sites at $(0.5,0)$ and $(0,0.5)$ are magnetic and host compensated magnetic moments.
As a difference to the model in Refs.~\cite{PhysRevB.108.224421,PhysRevLett.134.096703}, 
we include three $p$ orbitals at each atomic site; $p_x$, $p_y$, and $p_z$. As shown below, this allows us to explicitly model the 
intra-atomic contribution to the SOC among the $p$ states. 

The Hamiltonian is given by
\begin{align}\label{eq:Lieb-H}
    H &= \sum_{\langle i,j\rangle, \alpha}\!\! t^{(1)}_{ij} \Big(\hat{c}^{\dagger}_{i\alpha} \hat{c}_{j\alpha}  + \text{H.c.}\Big) + 
        \sum_{\langle\langle i,j\rangle\rangle, \alpha}\!\! t^{(2)}_{ij} \Big(\hat{c}^{\dagger}_{i\alpha} \hat{c}_{j\alpha}   + \text{H.c.}\Big)\nonumber \\
        &+J\sum_{i,\alpha,\beta}\mathbf{m}_i \cdot \hat{c}^{\dagger}_{i\alpha} \mathbf{S}_{\alpha \beta} \hat{c}_{i\beta} + \sum_{i,j,\alpha,\beta}\! \Big(\lambda_{i\alpha; j\beta} \hat{c}^{\dagger}_{i\alpha} \hat{c}_{j\beta} + \text{H.c.}\Big) ,
\end{align}
where $i$ and $j$ index orbitals located exclusively on magnetic atoms, and $\alpha$, $\beta$ denote spin indices.
The first two terms represent nearest-neighbor ($t_{1ij}$) and next-nearest-neighbor ($t_{2ij}$) hopping.
As there are three $p$ orbitals per site, we introduce the minimal distance between two orbitals $d_{ij}$ in order to weight the hopping strength as $t_{ij} = t_1/d_{ij}$, so that some hoppings are larger than others, that is, orbitals that are closer have a larger hopping strength. 
Additionally, for the next-nearest neighbor hopping, $t_{2ij} = t_{2a}/d_{ij}$ if the hopping path crosses the nonmagnetic atom, and $t_{2ij} = t_{2b}/d_{ij}$ otherwise.
The third term is the non-relativistic exchange interaction characteristic of altermagnetic materials~\cite{doi:10.1126/sciadv.aaz8809}, describing the coupling between the local magnetic moment $\mathbf{m}_i$ and the electron spin, where $\mathbf{S}$ is the vector of Pauli matrices. 
The final term in Eq.~\ref{eq:Lieb-H} represents the intra-atomic SOC, with the coupling constant between $p$-orbitals given by:
\begin{equation}
    \lambda_{i\alpha j\beta} = \bra{i\alpha} \mathbf{L}\cdot \mathbf{S} \ket{j\beta}. \label{eq:L.S}
\end{equation}
The non-zero matrix elements are~\cite{JAFFE1987399}:
\begin{align}
\begin{split}
    \lambda_{i\alpha j\beta} &= -i\lambda \quad (i\alpha = p_{x}^{\uparrow}, j\beta = p_y^{\uparrow}), \\
    \lambda_{i\alpha j\beta} &= i\lambda \quad (i\alpha = p_{x}^{\downarrow}, j\beta = p_y^{\downarrow}), \\
    \lambda_{i\alpha j\beta} &= \lambda \quad (i\alpha = p_{x}^{\uparrow}, j\beta = p_z^{\downarrow}), \\
    \lambda_{i\alpha j\beta} &= i\lambda \quad (i\alpha = p_{y}^{\uparrow}, j\beta = p_z^{\downarrow}).
\end{split}
\end{align}

We have considered the following numerical values of the model parameters 
are $a=5$~\AA, $J = t_1$, 
$t_{2a} = 2.5 t_1$, $t_{2b} = -1.5 t_1$,
with $t_1 = 0.5$~eV.
Additionally, a size of $0.3a$ was given to the $p$ orbitals in order to compute the minimal distance $d_{ij}$ between them.



For vanishing SOC, the band structure shows a 
compensated spin-splitting along $k_x$ and $k_y$ directions in the reciprocal lattice, which is a consequence of the $\{C_2 \| C_{4z}\}$ spin-symmetry that links opposite-spin sublattices (see first panel of Fig.~\ref{fig:model}b).
The non-relativistic spin splitting is generally 
of the order of $1$~eV, while bands remain spin degenerate along
$\Gamma-$M; these features are in agreement with  Refs.~\cite{PhysRevB.108.224421,PhysRevLett.134.096703}.
With finite $\lambda$ parameter, we observe a general modification of the band structure, accompanied by a change in the spin polarization, which is no longer restricted to purely up or down states but can instead take on intermediate values. 
In particular, the bands along the $\Gamma$-M direction become nondegenerate and develop a relativistic spin splitting and polarization.
For $\lambda = 0.1$~eV the size of the splitting 
is small but visible, of similar order to the effect of SOC
in many light-element materials. 
For $\lambda = 0.25$~eV the relativistic spin-splitting at $\Gamma$ is 
approximately $0.3$~eV,
while for $\lambda = 1$~eV it reaches $~1$~eV, on par with the non-relativistic splitting.
Therefore, we identify $\lambda < 0.1$~eV  with a  regime with weak relativistic effects, whereas $\lambda = 1$~eV marks a strong
relativistic regime (see last panels of Fig.~\ref{fig:model}b).

\subsection{Optical conductivity}

We now turn to studying the optical response of the model. 
We first derive symmetry predictions of both magnetic and spin
groups on the response 
tensor components of Eq.~\ref{eq:kubo}, 
and then present our numerical results.

\subsubsection{Symmetry predictions} 

The two-dimensional magnetic point group of the model shown in Fig.~\ref{fig:model} is $mm.1'$, comprising the following symmetry operations: the identity $\mathcal{I}$, two fold rotation $C_{2z}$, the mirrors $M_{x}$ and $M_{y}$ and their combination with time reversal $\mathcal{T}$.
In turn, when considering spin symmetries, additional operations emerge. These include 
$\{\mathcal{I} \| \mathcal{I}\}$, $\{\mathcal{I} \| C_{2z}\}$, $\{\mathcal{I} \| M_y\}$, 
$\{\mathcal{I} \| M_x\}$, and $\{C_{2z} \| C_{4z}\}$. The spatial components of these 
spin symmetries alone form the two-dimensional point group $4mm$.

We now examine the constraints imposed by magnetic and spin symmetries on the symmetric part of the optical conductivity tensor, $\sigma_S^{ab}(\omega)$.
Under magnetic symmetries, none of the operations relate the $x$ and $y$ axes;
this is evident from Fig.~\ref{fig:model}, 
where the orientation of the magnetic moments
clearly make the $x$ and $y$ axes non-equivalent.
Consequently, the independent components $\sigma^{xx}$ and $\sigma^{yy}$ are symmetry-allowed and unrelated.
In contrast, spin symmetries do impose constraints on the optical conductivity tensor. Specifically, from Eq.~\ref{eq:spin_dielectric}:
\begin{align}
    \sigma^{xx}_{S} &\xrightarrow{\{C_2\|C_{4z}\}} \sigma^{yy}_{S}.
\end{align}
By Neumann's principle, which requires that response tensors remain invariant under all crystal symmetries, we find 
\begin{align}\label{eq:ss-predic}
\sigma^{xx}_{S} &\doteq \sigma^{yy}_{S}.
\end{align}

\subsubsection{Numerical results}

Fig.~\ref{fig:model}b presents the calculated  optical conductivity components $\sigma^{xx}(\omega)$ and $\sigma^{yy}(\omega)$ (b.i–b.iv, lower panels) for increasing SOC strength $\lambda = 0,0.10,0.25,1$~eV.
We observe that for $\lambda=0$ the spin symmetry predictions of Eq.~\ref{eq:ss-predic} are exactly fulfilled, while for finite values deviations appear throughout the optical spectrum. For $\lambda=0.1$~eV and $\lambda=0.25$~eV only small differences are noticeable, whereas for $\lambda=1$~eV these become clearly visible, especially at low frequencies and in the peak structure around $\omega\sim4$~eV. Nevertheless, the overall trends of $\sigma^{xx}(\omega)$ and $\sigma^{yy}(\omega)$ show remarkable qualitative agreement even in the strong SOC limit of $\lambda = 1$~eV, despite the absence of any magnetic symmetry relating the two quantities.






%

To enable a more quantitative analysis, we consider 
the metric $\Delta$ defined by the normalized $L_2$ (Euclidian) norm between 
the diagonal components $\sigma^{xx}(\omega)$ and $\sigma^{yy}(\omega)$:
\begin{equation}
   \Delta = \dfrac{\sqrt{\displaystyle \int d \omega [\sigma^{xx}(\omega) - \sigma^{yy}(\omega)]^2} }{\sqrt{\displaystyle \int d \omega [\sigma^{xx}(\omega) + \sigma^{yy}(\omega)]^2}}. \label{eq:delta_metric}
\end{equation}
This metric is introduced as a measure of the overall anisotropy between the two diagonal components of the optical conductivity, and is directly related to the  linear dichroism, $\sigma^{xx}(\omega) - \sigma^{yy}(\omega)$.
In particular, 
$\Delta \simeq 0$ indicates isotropic behavior predicted by spin symmetries, namely no significant linear dichroism, 
while $\Delta \simeq 1$ implies a maximum possible difference, \textit{i.e.} high degree of linear dichroism in the optical response and thus, complete deviation from the SSG prediction.

The metric is displayed  in the bottom panel of 
Fig.~\ref{fig:model}c as a function of the 
SOC strength $\lambda$.
For comparison, we have also included a proxy quantity
$\Delta E_{\Gamma}(\lambda)$, defined as the difference between the two highest valence bands at $\Gamma$, which
roughly quantifies the effect of SOC on the eigenvalues.
The evolution of the metric $\Lambda$ indicates that, in the weak SOC regime ($\lambda < 0.25$~eV), the anisotropy remains below approximately 5~\%. Even in the strong SOC regime, approaching $\lambda = 1$~eV, the overall anisotropy stays below 15~\%. The model results therefore show that the two 
optical tensor components, related solely by spin symmetries, remain remarkably similar even in the strong SOC regime, despite the substantial modification of the band structure (Fig.~\ref{fig:model}b).
This evidences the intrinsic non-relativistic nature of the optical response,
which is therefore well described by spin symmetries even in the strong SOC regime.



\section{Validation through DFT calculations} 
\label{sec:dft}

Having established the behavior of a 
simple altermagnetic tight-binding model with intra-atomic SOC in the previous section, 
we now present results from more realistic DFT 
calculations for various magnetic materials
in which SOC is explicitly included in full fashion.

\subsection{Lieb lattice realization in actinide UCr$_2$Si$_2$C: large SOC limit}

\begin{figure*}
    \centering
    \includegraphics[width=\linewidth]{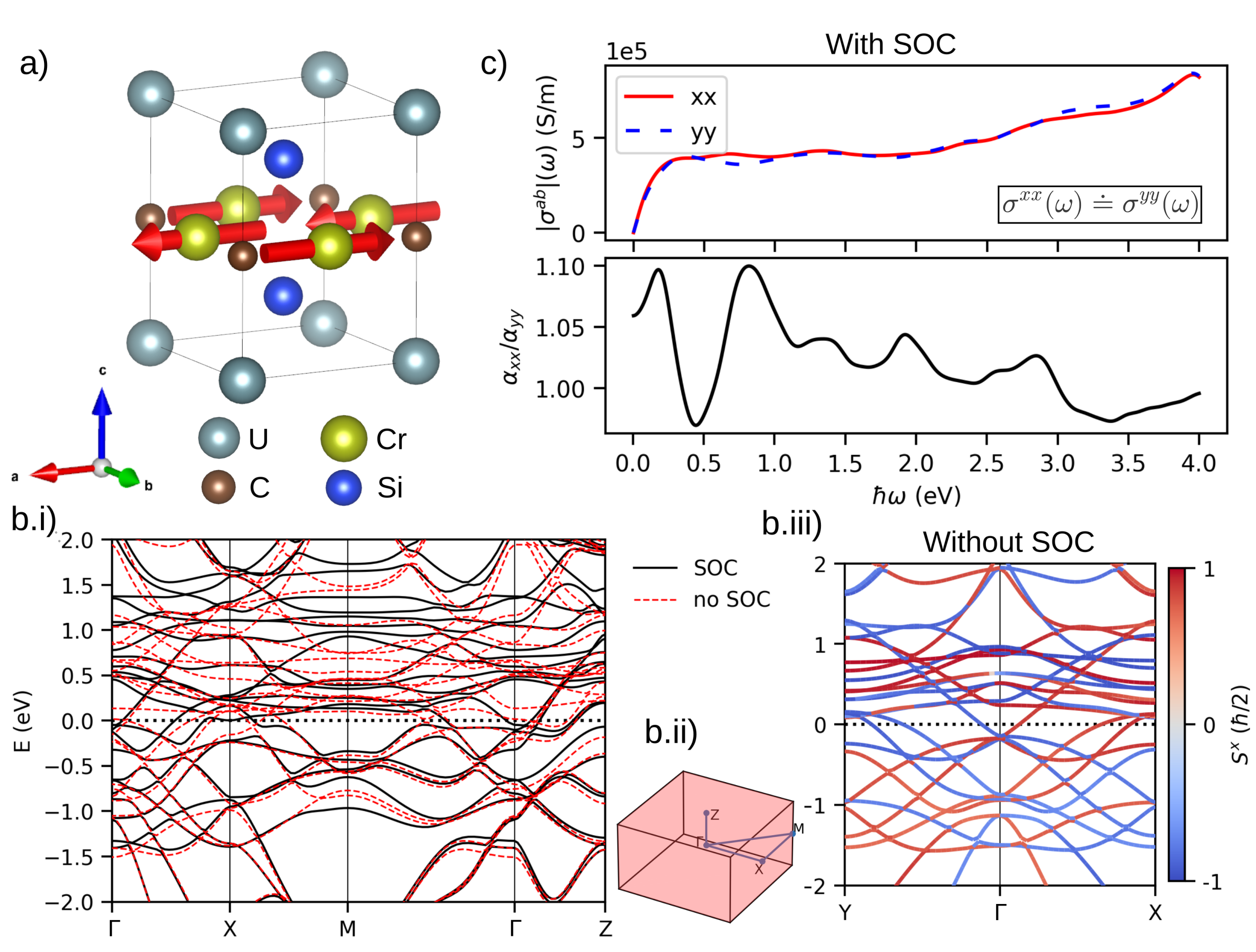}
    \caption{(a) Crystal structure of UCr$_2$Si$_2$C with the magnetic moments indicated by arrows. 
    The lattice vectors (in \AA) are $\mathbf{a_{1}} = (3.98,\,0,\,0)$, 
         $\mathbf{a_{2}} = (0,\,3.98,\,0)$, 
         $\mathbf{a_{3}} = (0,\,0,\,5.16)$.
    (b) Electronic band structure: 
            (b.i) band structure along a high-symmetry path; 
            (b.ii) Brillouin zone showing the k-path; 
            (b.iii) Altermagnetic spin-split bands along Y--$\Gamma$--X. The color scale encodes the spin polarization $S^x$. 
        (c) Top panel: Absolute value of the diagonal components of optical conductivity $\sigma^{xx}(\omega)$ and $\sigma^{yy}(\omega)$; the inset shows the expected relation between the components predicted by spin-symmetries. 
        Bottom panel: dichroic ratio $\alpha_{xx}(\omega)/\alpha_{yy}(\omega)$, where $\alpha_{aa}(\omega)$ is the absorption coefficient. }\label{fig:U}
\end{figure*}

As first material we consider is UCr$_2$Si$_2$C, which has attracted significant interest within the actinide community following experimental reports of long-range antiferromagnetic order.
Very recently, this compound has been identified as a $d$-wave altermagnetic candidate in a high-throughput materials study~\cite{sufyan2026}. Its crystal structure realizes the Lieb lattice model discussed in Sec.~\ref{sec:lieb}. It is also an ideal material platform to test the large-SOC regime analyzed there, due to the presence of heavy uranium atoms in the lattice structure.

UCr$_2$Si$_2$C crystallizes in the tetragonal structure shown in Fig.~\ref{fig:U}a.
Its parent nonmagnetic point group is $4/mmm$, but it hosts an unusual magnetic ground state: the Cr sublattice orders antiferromagnetically at high temperature ($T_{\mathrm N} > 300$~K), whereas the uranium moments remain magnetically disordered down to at least 2~K.
The magnetic ordering lowers the magnetic point group symmetry to $m'm'm$. However, when spin symmetries are taken into account, additional symmetry operations appear, most notably $\{C_2 || C_{4z}\}$. Consequently, the effective real-space point group restored by these spin symmetries is again $4/mmm$, corresponding to the full nonmagnetic point group.

Fig.~\ref{fig:U}b.i shows the metallic band structure of UCr$_2$Si$_2$C.
A comparison between calculations performed with and without SOC reveals its pronounced influence across several regions of the Brillouin zone (Fig.~\ref{fig:U}b.ii). In particular, near the $\Gamma$ and M points, SOC induces band splittings of nearly 0.5~eV.
In addition, clear signatures of d‑wave altermagnetism are visible in Fig.~\ref{fig:U}b.iii for band calculations without SOC, where 
the bands are colored according to their spin-polarization 
\begin{equation}\label{eq:spin-pol}
\boldsymbol{S}_{n}({\bf k})=\int \Psi^{*}_{{\bf k}n}({\bf r})
\hat{\boldsymbol{\sigma}} \Psi_{{\bf k}n}({\bf r})\textnormal{d}^{3}\mathbf{r},
\end{equation}
with $\hat{\boldsymbol{\sigma}}$ being the Pauli spin operator and $\Psi_{{\bf k}n}({\bf r})$ the Kohn–Sham spinor state associated with band $n$.
The spin symmetry operation $\{C_2 || C_{4z}\}$ relates opposite-spin sublattices, giving rise to the compensated altermagnetic spin polarization along the Y–$\Gamma$–X path.

We now turn to the optical absorption properties of UCr$_2$Si$_2$C.
In close analogy with the model discussed in Sec.~\ref{sec:lieb}, the spin symmetry $\{C_2 \| C_{4z}\}$ enforces the relation $\sigma^{xx}(\omega) \doteq \sigma^{yy}(\omega)$ between the diagonal components of the optical conductivity tensor, even though these components are not connected by the magnetic symmetries themselves. 
As shown in the top subplot of Fig.~\ref{fig:U}c, the calculated absorption components $\sigma^{xx}(\omega)$ and $\sigma^{yy}(\omega)$ in the presence of SOC are nearly identical across the entire optical frequency range, exhibiting a  very small residual anisotropy.
This small remaining anisotropy can be quantified by the ratio of absorption coefficients, often referred to as the dichroic ratio~\cite{ratio} $\alpha_{xx}(\omega) / \alpha_{yy}(\omega)$, where $\alpha_{aa}(\omega)$ is the absorption coefficient, related to the dielectric tensor via $\alpha_{aa}(\omega) = \frac{\omega}{c} \operatorname{Im} \sqrt{\varepsilon^{aa}(\omega)}$.
In contrast with the metric employed for the model calculations in Eq.~\ref{eq:delta_metric}, the dichroic ratio is an spectral magnitude.
Using the expression for the dielectric tensor $\varepsilon^{ab}(\omega)=i\sigma_{ab}(\omega)/\omega\varepsilon^0 + \delta_{ab}$, we evaluate the dichroic ratio and display it in the bottom subplot of Fig.~\ref{fig:U}c.
Despite the strong spin–orbit coupling associated with the uranium atoms, $\alpha_{xx}(\omega) / \alpha_{yy}(\omega)$ only deviates from the expected value 1.0 predicted by spin-symmetries by at most 10\%, reaching a value of approximately $1.1$ at low energies and remaining below $1.05$ for optical energies above $1$~eV, confirming that the two diagonal components are nearly equal across most of the spectrum.
For reference, the dichroic ratio ranges from $\sim 1.5$ in moderately anisotropic materials like ReSe$_2$~\cite{Zhong:21}, to $3$ in highly anisotropic systems like GeSe~\cite{10.1021/jacs.7b06314}.
%
Our results thus demonstrate that, even in systems containing heavy elements with large SOC, SSGs can still impose strong constraints on response tensors, provided the response coefficients are not dominated by relativistic effects.

%

\subsection{Birefringence in transition-metal fluoride RbMnF$_4$}


Up to this point, our discussion of the optical coefficients has focused on the diagonal components and their associated linear dichroism. While these coefficients primarily determine the optical absorption properties of a material, the off-diagonal components are of high interest too as they govern different set of optical phenomena. In this section, we focus on birefringence, namely the property of a material to exhibit different refractive indices depending on the polarization of the incident light. This effect is controlled by the symmetric part of the off-diagonal photoconductivity tensor components defined in Eq.~\ref{eq:sigma_A}. To illustrate this, we employ the altermagnetic candidate  RbMnF$_4$~\cite{sufyan2026}, for which SSGs predict birefringence effects that are not captured by the corresponding MSG analysis.

\begin{figure*}
    \centering
    \includegraphics[width=0.95\linewidth]{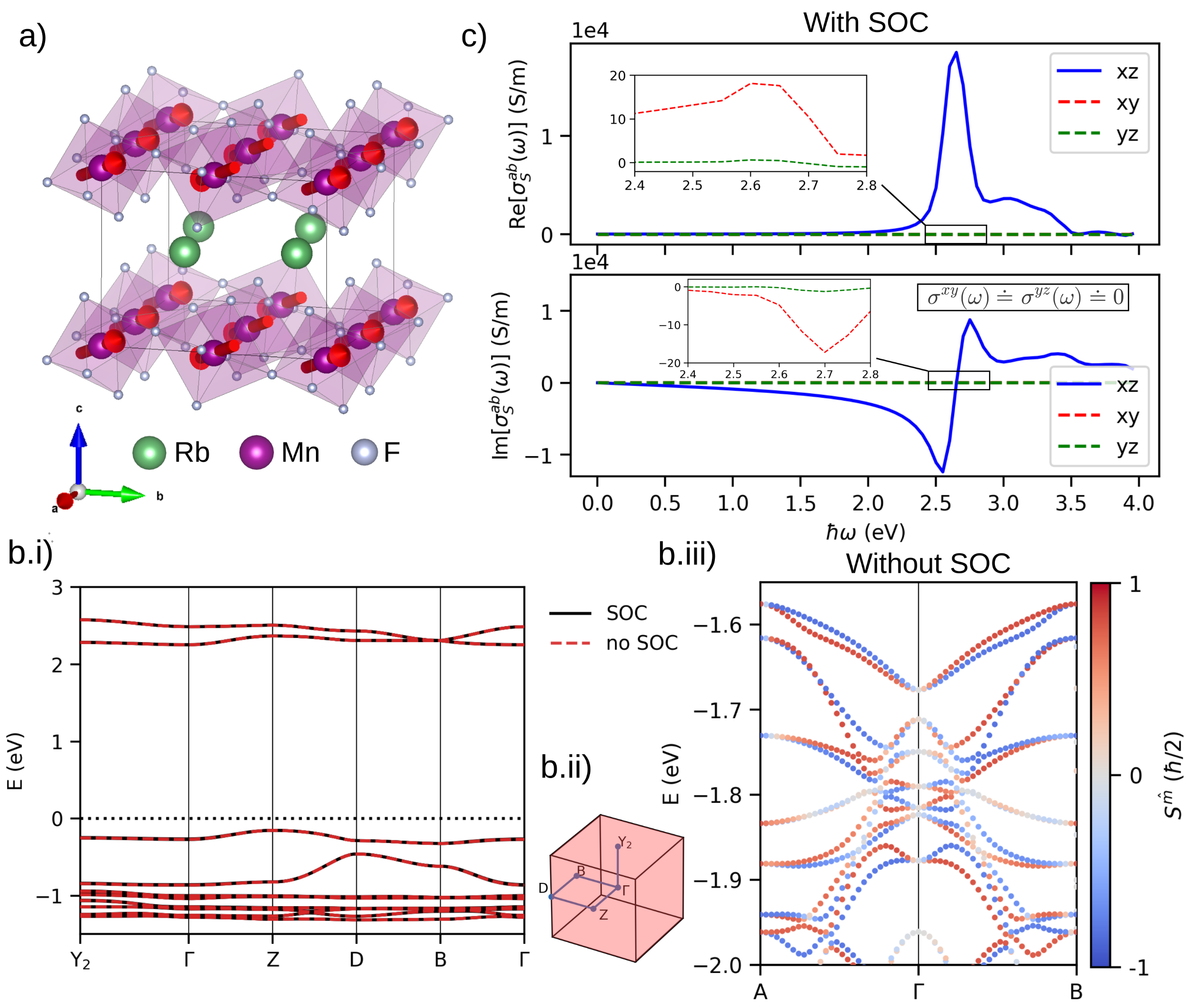}
    \caption{
         (a) Crystal structure of RbMnF$_4$ with the magnetic moments indicated by arrows. The lattice vectors (in \AA) are $\mathbf{a_{1}} = (7.95,\,0,\,0)$, 
         $\mathbf{a_{2}} = (0,\,7.91,\,0)$ and
         $\mathbf{a_{3}} = (-0.05,\,0,\,6.13)$.
         (b) Electronic band structure: 
            (b.i) band structure along the full high-symmetry path; 
            (b.ii) Brillouin zone showing the k-path; 
            (b.iii) zoom into the altermagnetic spin-split bands along A--$\Gamma$--B. The color scale encodes the spin polarization $S^{\hat{m}}$. 
         (c) Optical conductivity: top -- real part Re$[\sigma(\omega)]$ (main panel: off-diagonal components; inset: magnified $\sigma_{xy}$ and $\sigma_{yz}$ in the 2.4--2.8~eV range); bottom -- imaginary part Im$[\sigma(\omega)]$ with the same layout.}
    \label{fig:RbMnF4}
\end{figure*}

The unit cell of RbMnF$_4$ is shown in  Fig.~\ref{fig:RbMnF4}a.
The heaviest constituent element is Rb (atomic number $Z=37$), implying that SOC is non-negligible, though not dominant. The nonmagnetic point group of RbMnF$_4$ is $2/m$ (arising from the space group $P2_1/a$). The magnetic moments align along the direction $\bm{\hat{m}}=(1.97, 1.97, 1.03)$;
since this does not coincide with any crystallographic axis, the magnetic point group is greatly reduced to $-1.1$, containing only the identity and spatial inversion operations.
%
When spin symmetries are taken into account, additional operations emerge. The spin point group consists of the operations $\{\mathcal{I} \| \mathcal{I}\}$, $\{\mathcal{I} \| \mathcal{P}\}$, and crucially $\{C_2 \| M_y\}$. The latter is the spin operation that links opposite-spin sublattices, providing the fingerprint of altermagnetism. As a result, the effective real-space point group recovered from these spin symmetries is again $2/m$, which includes the real-space mirror operation $M_y$ 
that is not present in the magnetic point group.

Fig.~\ref{fig:RbMnF4}b shows the calculated band structure of RbMnF$_4$, which is predicted to be, at the DFT level with a semi-local functional (see Computational Details), a semiconductor with a wide gap of roughly 2.3~eV. The figure also displays the k‑space spin polarization.
Because the magnetic moments are collinear, any spin rotation about the $\hat{m}$ axis is a spin symmetry of the system (with $\hat{m}$ denoting the magnetization direction). Consequently, the spin operation $\{C_{2\hat{m}}\|\mathcal{I}\}$ relates the spin‑up and spin‑down projections along any direction perpendicular to $\hat{m}$ at the same $\mathbf{k}$ point. 
This symmetry forbids spin polarization along these directions so only $S^{\hat{m}}$ is plotted in Fig.~\ref{fig:RbMnF4}b. We note that small residual components along the in-plane directions can remain due to the presence of SOC.

Importantly, the zoomed region along A--$\Gamma$--B in Fig.~\ref{fig:RbMnF4}b.iii, with A = (0, 0.5, 0.5) and B = (0, -0.5, 0.5), reveals altermagnetic compensation: the spin operation $\{C_2 \| M_y\}$ maps points (0, k, k) onto (0, -k, k), thereby reversing the spin polarization. The same behavior occurs along the path C--$\Gamma$--D shown in the full band structure of panel (b.i), where C = (0.5, -0.5, 0) and D = (0.5, 0.5, 0). These momentum‑space patterns are the hallmark of d-wave altermagnetism.

Coming next to the optical tensor, MSGs do not impose  constraints on $\sigma^{ab}(\omega)$, which means that all components are allowed and  independent from each other.
When one considers SSGs, on the other hand, 
the mirror operation $M_y$ of the effective real-space point 
group requires that any off-diagonal tensor component with one 
$y$ index  vanish:
\begin{equation}\label{eq:ss-rb}
\sigma_{\mathrm{S}}^{xy}\doteq\sigma_{\mathrm{S}}^{yz}\doteq0.    
\end{equation}

The calculated optical conductivity, $\sigma_{\mathrm{S}}^{ab}(\omega)$, is displayed in Fig.~\ref{fig:RbMnF4}c. The top and bottom panels show the real and imaginary parts of the optical conductivity, respectively. In the energy range $2.4$--$2.8$~eV, $\sigma_{\mathrm{S}}^{xz}$ reaches values on the order of $10^4$~S/cm, whereas $\sigma_{\mathrm{S}}^{xy}$ and $\sigma_{\mathrm{S}}^{yz}$ remain at the $10^1$~S/cm scale. This strong suppression of the off-diagonal components containing the $y$ index is fully consistent with the spin-symmetry constraints [Eq.~\eqref{eq:ss-rb}], while the residual $xy$ and $yz$ signals originate from the weak SOC.

The presence of off-diagonal responses implies that the principal axes of the dielectric tensor are rotated away from the crystallographic axes, giving rise to birefringence. For each plane $(ab)$, the extinction angle $\phi^{ab}$, 
which measures the orientation of the principal dielectric axes relative to the crystallographic axis normal to the $ab$ plane, 
can be defined separately for the real and imaginary parts of the dielectric tensor~\cite{angle}: 
\begin{equation}
\tan(2\phi^{ab}_{\mathrm{Re}}) = \frac{2\,\mathrm{Re}(\varepsilon^{ab})}{\mathrm{Re}(\varepsilon^{aa})-\mathrm{Re}(\varepsilon^{bb})}, \qquad
\tan(2\phi^{ab}_{\mathrm{Im}}) = \frac{2\,\mathrm{Im}(\varepsilon^{ab})}{\mathrm{Im}(\varepsilon^{aa})-\mathrm{Im}(\varepsilon^{bb})}.
\end{equation}
The magnetic point group $-1.1$ leaves all components of $\varepsilon^{ab}(\omega)=\delta^{ab}+i\sigma^{ab}(\omega)/(\varepsilon^0\omega)$ independent, so birefringence and finite extinction angles would in principle be allowed in every plane. In turn, the monoclinic effective point group from spin-symmetry considerations restricts a non‑zero extinction angle to the $xz$ plane exclusively. Using the dielectric function obtained from the calculated optical conductivity, we find that in the $2.4$--$2.8$~eV range the real-part angle $\phi^{xz}_{\mathrm{Re}}$ exceeds $10^\circ$, while $\phi^{xy}_{\mathrm{Re}}$ and $\phi^{yz}_{\mathrm{Re}}$ stay around $0.5^\circ$. For the imaginary part, $\phi^{xz}_{\mathrm{Im}}$ reaches about $6^\circ$, with the other two angles near $0.3^\circ$. This confirms that the birefringence is overwhelmingly dominated by the $xz$ plane, a result that is not apparent 
from a simple inspection of the crystal structure or its magnetic point group.

\subsection{ScMnO$_3$: spin photoconductivity}

\begin{figure*}
    \centering
    \includegraphics[width=.95\linewidth]{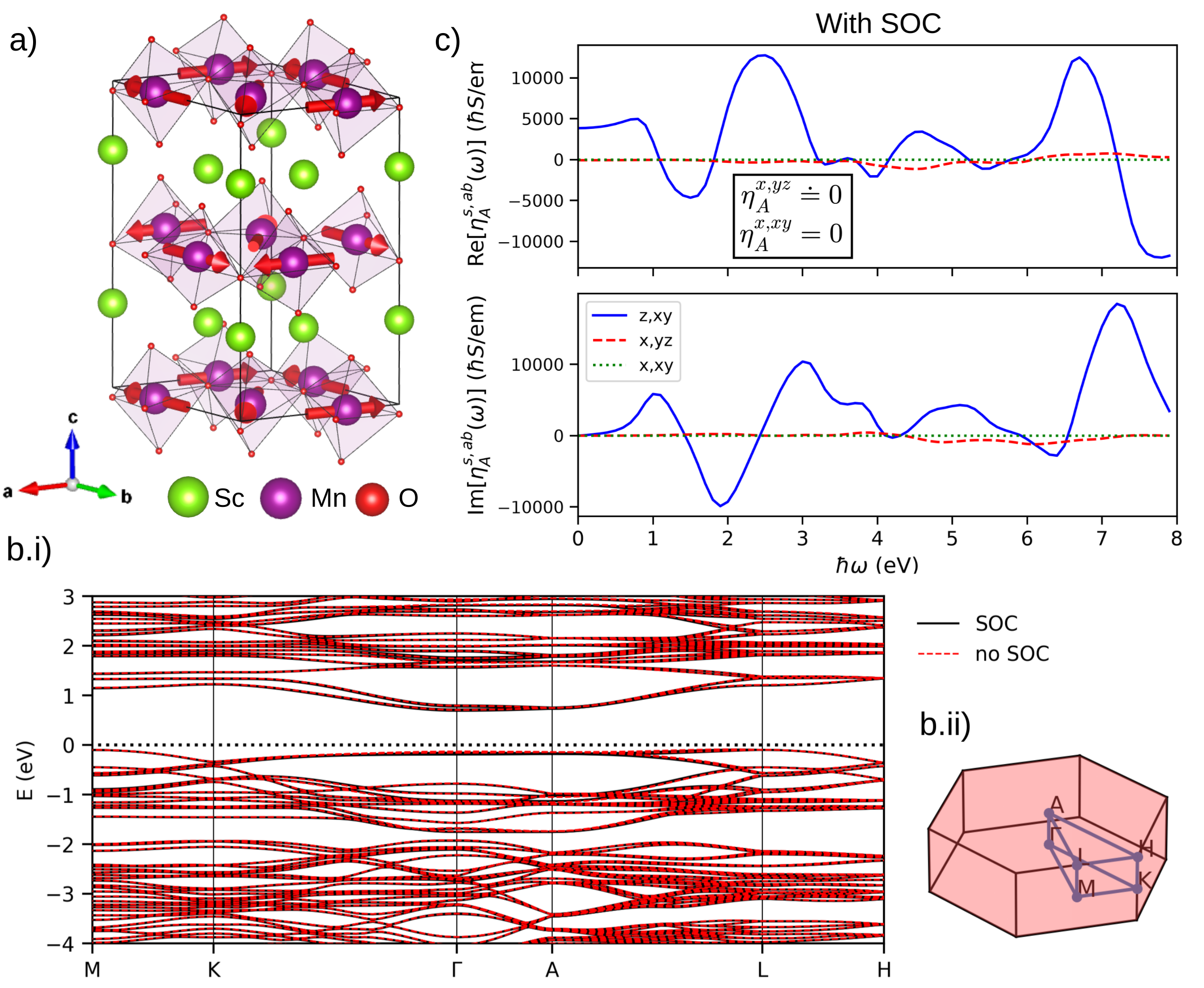}
    \caption{(a) Crystal structure of ScMnO$_3$ with the magnetic moments indicated by arrows.
    The lattice vectors (in \AA) are $\mathbf{a_{1}} = (5.84,\,0,\,0)$, 
         $\mathbf{a_{2}} = (-2.91,\,5.06,\,0)$, 
         $\mathbf{a_{3}} = (0,\,0,\,11.18)$.
    (b) Electronic band structure: (b.i) band structure along the full high-symmetry path; (b.ii) Brillouin zone showing the k-path
    (c) Spin Hall conductivity $\eta^{s,ab}_A(\omega)$: top -- real part; bottom -- imaginary part. The hierarchy follows the symmetry predictions: the dominant component, $\eta^{z,xy}_A(\omega)$, is allowed by both MSGs and SPGs; the weaker $\eta^{x,yz}_A(\omega)$ is allowed by MSGs but forbidden by SPGs; and $\eta^{x,xz}_A(\omega)$ vanishes identically, as it is forbidden by MSGs.}
    \label{fig:ScMnO3}
\end{figure*}

As a final example, we examine the \emph{spin} contribution to the optical response, with particular emphasis on the spin Hall conductivity and the way in which SSGs constrain this response coefficient.
Given their strong constraint on the spin Hall effect in collinear systems, we turn to a material belonging to the extended altermagnetic class introduced in Refs.~\cite{cheong2024altermagnetism, Cheong2025}, which generalizes the class to include non-collinear systems. As an illustrative example, we consider ScMnO$_3$, a coplanar magnet whose layered crystal structure is shown in Fig.~\ref{fig:ScMnO3}.

The magnetic point group is 
$6m'm'$, generated by the identity $\mathcal{I}$, a three-fold rotation $C_{3z}$, a 
two-fold rotation $C_{2z}$, and the combined operation $M_x\mathcal{T}$. The spin point 
group is minimal in this case, meaning that its operations consist of the magnetic 
symmetries applied simultaneously to both real and spin space, supplemented by a single 
spin-only operation. The latter arises from the coplanarity of the magnetic structure and takes the form $\{M_z \| \mathcal{I}\}$, which leaves the real-space structure invariant while reflecting the spin degrees of freedom.

The antisymmetric part of the spin Hall conductivity $\eta^{s,ab}_{A}$, which governs the intrinsic spin Hall effect, serves as a concrete observable through which the role of spin symmetries can be assessed.
Magnetic symmetries constrain $\eta_{A}^{s,ab}$ so that the only nonzero allowed components are $\eta_{A}^{x,yz}=\eta_{A}^{y,xz}$ and $\eta_{A}^{z,xy}$ (and $a \leftrightarrow b$). On top of these constraints, the spin-only operation $\{M_z \| \mathcal{I}\}$ 
introduces an additional restriction and applying Eq.~\ref{eq:spg_shc_a} yields
\begin{equation}\label{eq:eta-scmno}
\eta^{x,ab}_{A}\doteq\eta^{y,ab}_{A}\doteq0,
\end{equation}
so that the only allowed component is the z-spin-polarized element $\eta_{A}^{z,xy}$. 

Fig.~\ref{fig:ScMnO3}b shows the band structure of 
ScMnO$_3$, predicting a semiconducting ground state 
with a band gap of approximately 1~eV and negligible SOC effects.
The three relevant spin Hall coefficients are shown in
Fig.~\ref{fig:ScMnO3}c.
Notably, the calculated hierarchy closely follows the symmetry predictions: the dominant component, $\eta^{z,xy}_A(\omega)$, is more than an order of magnitude larger than the others and is allowed by both MSGs and SPGs; the much weaker $\eta^{x,yz}_A(\omega)$ is allowed by MSGs but forbidden by SPGs, and therefore arises solely from relativistic effects; and $\eta^{x,xz}_A(\omega)$ vanishes identically because it is forbidden by MSGs.
In practice, this means that the generated spin currents can carry only a spin polarization perpendicular to the magnetization plane, in full agreement with the spin-symmetry prediction of Eq.~\ref{eq:eta-scmno}.
This result, together with the previous cases examined throughout this section, clearly highlights the usefulness of SSGs in predicting a broad range of optical properties in magnetic materials.

\section{Conclusions and Outlook}

In this work, we have shown that SSGs provide a natural and powerful framework for analyzing the optical response of magnetic materials. While conventional MSG analysis captures the relativistic symmetry of a magnetic crystal, it can miss robust constraints that survive in observables governed mainly by the nonrelativistic electronic structure. By decoupling spin and real-space operations, SSGs generate approximate symmetries that can impose additional relations among optical coefficients, even in the presence of finite SOC. This makes them especially useful for altermagnets and related magnetic systems, where opposite-spin sublattices are connected by nontrivial spin symmetries.

For charge optical responses, SSGs induce effective spatial symmetries whose associated point groups provide an accurate guide to the allowed structure of the photoconductivity tensor. We illustrate this effect in an altermagnetic Lieb lattice model~\cite{PhysRevB.108.224421,PhysRevLett.134.096703}, which offers a controlled platform for tuning the strength of SOC and systematically exploring its consequences on the bandstructure and optical properties. Although the system is \textit{a priori} magnetically anisotropic, as the magnetic moments are aligned along the crystallographic $x$ axis, 
we nevertheless observe only weak linear dichroism even for moderate SOC strengths, a  behavior that reflects the  relation $\sigma^{xx} \doteq \sigma^{yy}$ inherited from the underlying spin symmetry. This exemplifies how the optical response can retain a marked nonrelativistic character well beyond the limit of vanishing SOC. 

We corroborate this principle through a systematic
\textit{ab-initio} analysis in realistic Lieb-lattice altermagnetic
candidates. 
We first examine the actinide UCr$_2$Si$_2$C,  where despite pronounced relativistic band splittings, its optical absorption remains nearly isotropic, in full agreement with our model predictions. We then focus on the transition-metal fluoride  RbMnF$_4$, where spin symmetries  suppress 
selected \emph{off-diagonal} optical conductivity components. Notably, this confines the birefringent response to a single plane, 
revealing a distinctive symmetry fingerprint  absent
from the magnetic-group analysis.

To demonstrate that these principles are not limited to the optical charge response of altermagnets, in the last part we extend the analysis to the spin Hall response of the coplanar noncollinear antiferromagnet ScMnO$_3$. 
Here, a spin-only symmetry operation imposes an additional constraint beyond magnetic-group symmetry, suppressing the in-plane spin-polarized conductivity components and leaving the out-of-plane response dominant. This result establishes that the predictive power of spin-group symmetries extends beyond standard altermagnetic systems and applies equally to spin-current response functions, opening a broader route to symmetry-guided design in spintronics.

\begin{table}[t]
\caption{\label{table4wann}
Summary of relevant predictions 
of MSGs  and SSGs 
on charge and spin 
response coefficients for the various systems 
analyzed throughout our work.
The symbol $\doteq$  denotes 
``non-relativistic'' constraints by 
spin groups.
    }
    \begin{ruledtabular}
 \begin{tabular}{  c | c |  c |  c | c  }
  & \multicolumn{2}{c|}{Magnetic symmetries} &  \multicolumn{2}{c}{Spin symmetries}  \\
  \hline
  & Point group & Constraint on $\sigma^{ab}$ or $\eta^{s,ab}$ & Relevant symmetry & Constraint on $\sigma^{ab}$ or $\eta^{s,ab}$\\
    \hline
Lieb lattice &  mm.1' & $\sigma^{xx}$ and $\sigma^{yy}$ unrelated & $\{C_{2z} \| C_{4z}\}$ & $\sigma^{xx} \doteq \sigma^{yy}$ \\ 
UCr$_2$Si$_2$C & m'm'm & $\sigma^{xx}$ and $\sigma^{yy}$ unrelated & $\{C_{2z} \| C_{4z}\}$ & $\sigma^{xx} \doteq \sigma^{yy}$ \\ 
RbMnF$_4$ & -1.1 & Unrestricted & $\{C_2\|M_y\}$ & $\sigma^{xy}\doteq\sigma^{yz}\doteq0$ \\ 
ScMnO$_3$ & 6m'm' & $\eta^{x,yz}$ allowed & $\{M_z\|\mathcal{I}\}$ & $\eta^{x,yz}\doteq0$ \\ 
\end{tabular}
\end{ruledtabular}
\end{table}

%
%

The present work has focused on the symmetric optical response and spin-current response channels, whose principal symmetry predictions are concisely summarized in Table~\ref{table4wann} for ease of reference, while deliberately excluding the antisymmetric off-diagonal conductivity, which underlies circular magnetodichroic phenomena~\cite{PhysRevLett.132.176701,q8ym-l2zt,10.1021/acs.nanolett.5c03647}, including the anomalous Hall effect, the magneto-optical Kerr effect, and Faraday rotation. 
For the mainly collinear antiferromagnets considered here, this contribution vanishes in the absence of spin-orbit coupling, so spin-group symmetry predicts $\sigma_A^{ij} \doteq 0$ without yielding qualitatively new constraints. A more promising setting for nontrivial spin-symmetry predictions in this channel is provided by noncollinear antiferromagnets, where a finite spin chirality can act as an effective magnetic field and enable circular dichroic responses even without spin-orbit coupling~\cite{Feng2020,Farhang2026}.

Taken together, our results establish spin symmetries as an organizing principle for optical phenomena in magnetic materials. They explain why response coefficients that appear unrelated from the perspective of magnetic symmetries can nevertheless exhibit nearly identical values or remain strongly suppressed in realistic calculations. 
These results
are likely to be relevant to the growing body of recent work on optical responses in altermagnets~\cite{weber2026opticalexcitationspinpolarization,Weber2025,PhysRevB.110.024425,vila2024orbitalspinlockingopticalsignatures,Sunko2026,KIMEL2024172039,463y-q7lt}.
More broadly, our work demonstrates that hidden nonrelativistic symmetries can leave clear and measurable fingerprints in both charge and spin optical responses, providing a powerful framework for interpreting optical experiments and guiding the search for magnetic materials with attractive optical functionalities.

\begin{acknowledgments}
We thank I. Souza  for clarifying discussions.
J. S., E. B.-S. and J. I.-A. acknowledge the financial support of the European Union’s Horizon 2020 research and innovation programme under the European Research Council (ERC) grant agreement No. 946629 StG PhotoNow. 
D. H.-P. is grateful for funding from the Diputaci\'on Foral de Gipuzkoa through Grants 2023-FELL-000002-01, 2024-FELL-000009-01, 2025-FELL-000004-01 and Red
Guipuzcoana R\&D Project TRAFIC (Project 2025-CIE4-000036-01).
D. H.-P. and J. I.-A. are also grateful from support of the Spanish {MICIU/AEI
/10.13039/501100011033} and ERDF, UE through Project No. PID2023-147324NA-I00 and from IKUR Strategy, Quantum Technologies 2025 project \textit{M-Twist} from the Department of Science, Universities and Innovation of the Basque Government.
\end{acknowledgments}

\appendix 

\section{Charge and spin responses}

\subsection{{Optical conductivity}} 

The optical conductivity within the independent-particle approximation is commonly formulated using the Kubo-Greenwood expression~\cite{doi:10.1143/JPSJ.12.570, Greenwood_1958}
\begin{equation}
\sigma^{ab}(\omega) = \frac{ie^{2}}{\hbar V N_{\mathbf{k}}} \sum_{\mathbf{k}} \sum_{n,m \neq n} f_{mn} A_{nm}^{a}(\mathbf{k}) A_{mn}^{b}(\mathbf{k}) 
\dfrac{\varepsilon^{m\mathbf{k}} - \varepsilon^{n\mathbf{k}}}{\varepsilon^{m\mathbf{k}} - \varepsilon^{n\mathbf{k}} - (\hbar \omega + i \eta)}, \label{eq:kubo}
\end{equation}
where $V$ is the crystal volume, $N_\mathbf{k}$ is the number of \textbf{k}-points used for the Brillouin zone sampling, $\hbar \omega$ is the photon energy, $\eta$ is an infinitesimal positive quantity and the indices $a,b$ denote Cartesian coordinates.
The quantity $A_{nm}^{a}(\mathbf{k}) := i\langle u_{n\mathbf{k}}| \partial_{k_a} u_{m\mathbf{k}}\rangle$ represents the Berry connection between cell-periodic Bloch states.
The difference in the occupation factors, $f_{mn} = f_m - f_n$ with $f_n := f(\varepsilon^{n\mathbf{k}})$ ensures that only transitions between occupied and unoccupied states contribute. The occupations are given by the Fermi-Dirac distribution,
\begin{equation}
    f(E) = \dfrac{1}{1 + \exp[(E -\mu)/k_\textnormal{B}T]},
\end{equation}
evaluated at energy $E$, temperature $T$ and chemical potential $\mu$ for the global equilibrium state ($k_\textnormal{B}$ represents the Boltzmann constant).
%

Using the Sokhotski-Plemelj formula on the real line
\begin{equation}
   \lim_{\eta \to 0}  \dfrac{1}{x-i\eta} =  \mathcal{P}\left(\frac{1}{x}\right) + i\pi \delta(x) \label{eq:SP_formula}
\end{equation}
we can decompose the optical conductivity into two contributions.
The real part, which contains the delta function, enforces energy conservation and corresponds to real interband transitions - true absorption or emission processes.
In contrast, the imaginary part describes off-resonant virtual processes and can therefore produce a finite response even for photon energies below the band gap.

\subsection{Spin conductivity} 

The spin conductivity within the independent-particle framework can also be described with a similar Kubo-Greenwood formula~\cite{Ryoo2019}
\begin{equation}
    \eta_s^{ab}(\omega) = \frac{e}{\hbar V N_\mathbf{k}} \sum_{\mathbf{k}}\sum_{n, m\neq n}\dfrac{ f_{nm} \text{Im}\left[ j^{s,a}_{mn} (\mathbf{k})v_{nm}^{b} (\mathbf{k})\right] }{(\varepsilon^{n\mathbf{k}} -\varepsilon^{m\mathbf{k}})^2 - (\hbar \omega + i\eta)^2}, \label{eq:shc}
\end{equation}
where the spin current operator \cite{Sinova2015} is defined as 
\begin{equation}
    \hat{j}^{s,a} := \frac{1}{2}\{\hat{v}^{a}, \hat{S}^{s}\},
\end{equation}
with $\hat{\mathbf{S}} = \hbar \bm{\sigma}/2$ the spin operator, $\bm{\sigma} = (\sigma_x, \sigma_y, \sigma_z)$ a vector whose components are the Pauli matrices, $\hat{\mathbf{v}}$ the velocity operator and $\{\cdot, \cdot\}$ the anticommutator.
%
Using the identity $v_{mn}^b(\mathbf{k}) = (i/\hbar) (\varepsilon^{m\mathbf{k}} - \varepsilon^{n\mathbf{k}}) A^b_{mn}(\mathbf{k})$ for $n\neq m$, we can rewrite Eq. \eqref{eq:shc} as 
\begin{equation}
    \eta_s^{ab}(\omega) = -\frac{e}{\hbar V N_\mathbf{k}} \sum_{\mathbf{k}}\sum_{n, m\neq n} \dfrac{f_{mn} \text{Re}\left[ j^{s,a}_{mn} (\mathbf{k})A_{nm}^{b} (\mathbf{k})\right] }{\varepsilon^{m\mathbf{k}} -\varepsilon^{n\mathbf{k}} - (\hbar \omega + i\eta)}, \label{eq:shc_2}
\end{equation}
which is the form implemented in WannierBerri \cite{Tsirkin2021}.

\section{Methods}

\subsection{Wannier interpolation}

For the computation of response tensors, we employed Wannier interpolation. This technique exploits the localized real-space character of Wannier functions to construct an accurate tight-binding Hamiltonian that can be efficiently evaluated at any point in momentum space.
The Wannier interpolation workflow begins with the Bloch eigenstates, \( \ket{\psi_{n\mathbf{k}}}  = |u_{n\mathbf{k}} \rangle {\rm e}^{i \mathbf{k} \cdot \mathbf{r}}\), which are obtained from a first-principles density functional theory (DFT) calculation on a coarse \(\mathbf{k}\)-point grid.
These states are transformed into a set of Wannier functions, \( \ket{n\mathbf{R}} \), by means of a unitary rotation of the Bloch manifold followed by a Fourier transform
\begin{equation}
\ket{n\mathbf{R}} = \frac{V}{(2\pi)^3} \int_{\text{BZ}} d\mathbf{k} \, e^{-i\mathbf{k}\cdot\mathbf{R}} \sum_{m} U_{mn}^{(\mathbf{k})} \ket{\psi_{m\mathbf{k}}},
\label{eq:bloch_to_wannier}
\end{equation}
where \( \mathbf{R} \) is a lattice vector, \( V \) is the unit-cell volume, and \( U^{(\mathbf{k})} \) is a unitary matrix that mixes the Bloch states at wavevector \(\mathbf{k}\).
A standard next step involves to exploit the gauge freedom in \( U^{(\mathbf{k})} \) to minimize the spatial spread of the resulting Wannier functions, producing so-called maximally localized Wannier functions (MLWFs)~\cite{PhysRevB.56.12847}. 
However, the localization can potentially break intrinsic crystal symmetries.
Although methods exist to explicitly preserve symmetries during the localization process~\cite{PhysRevB.87.235109}, 
an alternative practical approach is to avoid the localization altogether when the initial Wannier functions are already sufficiently localized~\cite{PhysRevB.87.235109}, thus preserving the symmetry of the states without further intervention.

In the resulting Wannier basis, the real-space Hamiltonian matrix elements \( H_{mn}(\mathbf{R}) = \bra{m\mathbf{0}} \hat{H} \ket{n\mathbf{R}} \) are computed. The Hamiltonian at an arbitrary wavevector \(\mathbf{k}\) is then reconstructed using a Fourier transform
\begin{equation}
H_{mn}(\mathbf{k}) = \sum_{\mathbf{R}} e^{i\mathbf{k}\cdot\mathbf{R}} H_{mn}(\mathbf{R}).
\label{eq:hamiltonian_fourier}
\end{equation}
This Fourier-based interpolation scheme allows for the efficient calculation of band energies \( \varepsilon^n(\mathbf{k}) \) and Bloch states by diagonalizing \( H(\mathbf{k}) \) on an arbitrarily dense \(\mathbf{k}\)-point grid. The same interpolation strategy applies to any operator expressed in the Wannier basis, including the position
operator, 
enabling the efficient evaluation of all quantities required for computing response tensors on arbitrarily dense \textbf{k}-point grids.


%
%
%

\subsection{Computational details}

\subparagraph{\small Tight-binding model.} The Lieb lattice 
tight-binding model was built using the \texttt{PythTB} package~\cite{Cole_Python_Tight_Binding_2025}.
The calculation of the response properties was subsequently carried out with the \texttt{WannierBerri} code~\cite{Tsirkin2021}, employing 
a grid of $100\times100$ and a fixed smearing parameter of $0.2$ eV.
The matrix elements of the position operator, $\hat{\mathbf{r}}$, were evaluated within the tight-binding approximation, in which the position operator is assumed to be diagonal in the Wannier basis and nonzero only for Wannier functions located at the same lattice site, \textit{i.e.}
$   \bra{n\mathbf{0}} \hat{\mathbf{r}} \ket{m\mathbf{R}} = \boldsymbol{\tau}_{n} \delta_{\mathbf{0}\mathbf{R}} \delta_{nm}$.
This approximation is consistent with the localized nature of the Wannier functions and is commonly employed in tight-binding implementations of linear and nonlinear response theories.


\subparagraph{\small UCr$_2$Si$_2$C.} Fully relativistic density functional theory (DFT) calculations were performed using the \texttt{Vienna Ab initio Simulation Package} (VASP)~\cite{PhysRevB.47.558, PhysRevB.54.11169}. 
Projector augmented-wave (PAW) pseudopotentials~\cite{PhysRevB.59.1758} with the Perdew-Burke-Ernzerhof (PBE) exchange-correlation functional~\cite{Perdew1996} were employed.
A plane-wave energy cutoff of $600$ eV and a $12\times12\times12$ $\mathbf{k}$-point grid were used. SOC was included self-consistently.
An on-site Hubbard $U$ correction of $3.5$ eV was applied to the Cr $d$ states following Dudarev’s approach~\cite{PhysRevB.57.1505, mmdm-hrj4}.
The magnetic configuration was initialized with the experimental Cr moment of $\sim 0.66\,\mu_{B}$~\cite{Gallego:ks5530, Gallego:ks5532}, but the self-consistent calculation yielded a larger moment of $\sim 2.5\,\mu_{B}$; this does not affect the magnetic symmetries, which are the relevant aspect for our work.

\subparagraph{\small RbMnF\textsubscript{4}.} The electronic structure of RbMnF$_4$ was obtained following the same DFT methodology as for UCr$_2$Si$_2$C, with a plane-wave energy cutoff of $500$ eV, a $\mathbf{k}$-point grid of $8\times 8 \times 8$, and an on-site Hubbard $U$ of $3$ eV applied to the Mn $d$ states~\cite{PhysRevB.57.1505}.
A Wannier Hamiltonian was subsequently constructed using \texttt{Wannier90}~\cite{pizzi_wannier90_2019} by projecting onto Mn $d$ and F $p$ orbitals to create a model comprising $136$ Wannier functions.
The inner and outer energy windows were both set to $(-8.35, 3.65)$ eV to isolate the relevant bands manifolds.
To preserve the symmetries of the underlying DFT calculations, the localization procedure was performed with zero minimization iterations. Despite the absence of maximal localization, the resulting Wannier functions remain well localized, with a maximum individual spread of $0.75$~\AA\ and a total spread of $87.3$~\AA\ across all 136 functions.

\subparagraph{\small ScMnO\textsubscript{3}.} For ScMnO$_3$, the electronic structure was obtained using the same DFT approach, with a plane-wave cutoff energy of $500$ eV and an $8 \times 8 \times 8$ $\mathbf{k}$-point grid.
The Wannier representation was generated using Wannier90~\cite{pizzi_wannier90_2019} using projections from Mn $d$ and Sc $p$ and $s$ orbitals, resulting in a total of 204 Wannier functions.
The energy windows were chosen such that all bands were included at the lower-energy end, while the upper bounds of the inner and outer windows were set to $7$ eV and $8$ eV, respectively.
As in the case of RbMnF$_4$, the number of localization iterations was set to zero to retain the crystal symmetries inherited from the DFT calculations.
The Wannier functions exhibit good localization properties, with a maximum spread of $1.3$~\AA\ and a cumulative spread of $228.6$~\AA\ for the full set of 204 functions.


\bibliography{biblio}

@article{Cheong2025,
author={Cheong, Sang-Wook and Huang, Fei-Ting},
title={Altermagnetism classification},
journal={npj Quantum Materials},
year={2025},
month={Apr},
day={12},
volume={10},
number={1},
pages={38},
doi={10.1038/s41535-025-00756-5},
url={https://doi.org/10.1038/s41535-025-00756-5}
}

@article{PhysRevB.108.224421,
  title = {Two-dimensional altermagnets: Superconductivity in a minimal microscopic model},
  author = {Brekke, Bj\o{}rnulf and Brataas, Arne and Sudb\o{}, Asle},
  journal = {Phys. Rev. B},
  volume = {108},
  issue = {22},
  pages = {224421},
  numpages = {11},
  year = {2023},
  month = {Dec},
  publisher = {American Physical Society},
  doi = {10.1103/PhysRevB.108.224421},
  url = {https://link.aps.org/doi/10.1103/PhysRevB.108.224421}
}

@article{PhysRevX.12.040501,
  title = {Emerging Research Landscape of Altermagnetism},
  author = {\ifmmode \check{S}\else \v{S}\fi{}mejkal, Libor and Sinova, Jairo and Jungwirth, Tomas},
  journal = {Phys. Rev. X},
  volume = {12},
  issue = {4},
  pages = {040501},
  numpages = {27},
  year = {2022},
  month = {Dec},
  publisher = {American Physical Society},
  doi = {10.1103/PhysRevX.12.040501},
}

@article{PhysRevX.12.031042,
  title = {Beyond Conventional Ferromagnetism and Antiferromagnetism: A Phase with Nonrelativistic Spin and Crystal Rotation Symmetry},
  author = {\ifmmode \check{S}\else \v{S}\fi{}mejkal, Libor and Sinova, Jairo and Jungwirth, Tomas},
  journal = {Phys. Rev. X},
  volume = {12},
  issue = {3},
  pages = {031042},
  numpages = {16},
  year = {2022},
  month = {Sep},
  publisher = {American Physical Society},
  doi = {10.1103/PhysRevX.12.031042},
}

@Article{Ma2021,
author={Ma, Hai-Yang
and Hu, Mengli
and Li, Nana
and Liu, Jianpeng
and Yao, Wang
and Jia, Jin-Feng
and Liu, Junwei},
title={Multifunctional antiferromagnetic materials with giant piezomagnetism and noncollinear spin current},
journal={Nature Communications},
year={2021},
month={May},
day={14},
volume={12},
number={1},
pages={2846},
issn={2041-1723},
doi={10.1038/s41467-021-23127-7},
url={https://doi.org/10.1038/s41467-021-23127-7}
}

@article{PhysRevB.102.014422,
  title = {Giant momentum-dependent spin splitting in centrosymmetric low-$Z$ antiferromagnets},
  author = {Yuan, Lin-Ding and Wang, Zhi and Luo, Jun-Wei and Rashba, Emmanuel I. and Zunger, Alex},
  journal = {Phys. Rev. B},
  volume = {102},
  issue = {1},
  pages = {014422},
  numpages = {13},
  year = {2020},
  month = {Jul},
  publisher = {American Physical Society},
  doi = {10.1103/PhysRevB.102.014422},
 }

@article{Hayami_2019,
   title={Momentum-Dependent Spin Splitting by Collinear Antiferromagnetic Ordering},
   volume={88},
   ISSN={1347-4073},
     DOI={10.7566/jpsj.88.123702},
   number={12},
   journal={Journal of the Physical Society of Japan},
   publisher={Physical Society of Japan},
   author={Hayami, Satoru and Yanagi, Yuki and Kusunose, Hiroaki},
   year={2019},
   month=dec }

@article{PhysRevLett.134.096703,
  title = {Mirror {C}hern Bands and {W}eyl Nodal Loops in Altermagnets},
  author = {Antonenko, Daniil S. and Fernandes, Rafael M. and Venderbos, J\"orn W. F.},
  journal = {Phys. Rev. Lett.},
  volume = {134},
  issue = {9},
  pages = {096703},
  numpages = {7},
  year = {2025},
  month = {Mar},
  publisher = {American Physical Society},
  doi = {10.1103/PhysRevLett.134.096703},
  url = {https://link.aps.org/doi/10.1103/PhysRevLett.134.096703}
}

@article{
doi:10.1126/sciadv.aaz8809,
author = {Libor Šmejkal  and Rafael González-Hernández  and T. Jungwirth  and J. Sinova },
title = {Crystal time-reversal symmetry breaking and spontaneous {H}all effect in collinear antiferromagnets},
journal = {Science Advances},
volume = {6},
number = {23},
pages = {eaaz8809},
year = {2020},
doi = {10.1126/sciadv.aaz8809},
url = {https://www.science.org/doi/abs/10.1126/sciadv.aaz8809}}

@article{JAFFE1987399,
title = {Inclusion of spin-orbit coupling into tight binding bandstructure calculations for bulk and superlattice semiconductors},
journal = {Solid State Communications},
volume = {62},
number = {6},
pages = {399-402},
year = {1987},
issn = {0038-1098},
doi = {https://doi.org/10.1016/0038-1098(87)91042-8},
url = {https://www.sciencedirect.com/science/article/pii/0038109887910428},
author = {M.D. Jaffe and J. Singh}
}

@article{PhysRevX.14.031038,
  title = {Enumeration and Representation Theory of Spin Space Groups},
  author = {Chen, Xiaobing and Ren, Jun and Zhu, Yanzhou and Yu, Yutong and Zhang, Ao and Liu, Pengfei and Li, Jiayu and Liu, Yuntian and Li, Caiheng and Liu, Qihang},
  journal = {Phys. Rev. X},
  volume = {14},
  issue = {3},
  pages = {031038},
  numpages = {33},
  year = {2024},
  month = {Aug},
  publisher = {American Physical Society},
  doi = {10.1103/PhysRevX.14.031038},
  url = {https://link.aps.org/doi/10.1103/PhysRevX.14.031038}
}

@article{doi:10.1098/rspa.1966.0211,
author = {Brinkman, W. F.  and Elliott, Roger James},
title = {Theory of spin-space groups},
journal = {Proc. R. Soc. Lond. A Math. Phys. Sci.},
volume = {294},
number = {1438},
pages = {343-358},
year = {1966},
doi = {10.1098/rspa.1966.0211},
url = {https://royalsocietypublishing.org/doi/abs/10.1098/rspa.1966.0211}
}

@misc{hellenes2024pwavemagnets,
      title={P-wave magnets}, 
      author={Anna Birk Hellenes and Tomáš Jungwirth and Rodrigo Jaeschke-Ubiergo and Atasi Chakraborty and Jairo Sinova and Libor Šmejkal},
      year={2024},
      eprint={2309.01607},
      archivePrefix={arXiv},
      primaryClass={cond-mat.mes-hall},
      url={https://arxiv.org/abs/2309.01607}, 
}

@article{tdrm-twnt,
  title = {Optically controllable spin polarization in two-dimensional altermagnets via robust spin-momentum locking excitons},
  author = {Sun, Jiuyu and Han, Jinzhe and Du, Yongping and Kan, Erjun},
  journal = {Phys. Rev. B},
  volume = {112},
  issue = {24},
  pages = {245417},
  numpages = {12},
  year = {2025},
  month = {Dec},
  publisher = {American Physical Society},
  doi = {10.1103/tdrm-twnt},
  url = {https://link.aps.org/doi/10.1103/tdrm-twnt}
}

@article{cqn4-lljy,
  title = {Valley-selective linear dichroism and excitonic effects in Lieb-lattice altermagnets},
  author = {Wang, Haonan and Xu, Xilong and Li, Du and Yang, Li},
  journal = {Phys. Rev. B},
  volume = {113},
  issue = {11},
  pages = {115408},
  numpages = {8},
  year = {2026},
  month = {Mar},
  publisher = {American Physical Society},
  doi = {10.1103/cqn4-lljy},
  url = {https://link.aps.org/doi/10.1103/cqn4-lljy}
}

@article{zn7r-k1xd,
  title = {Symmetry Classification for Alternating Excitons in Two-Dimensional Altermagnets},
  author = {Cao, Jiayu David and Denisov, Konstantin S. and Liu, Yuntian and \ifmmode \check{Z}\else \v{Z}\fi{}uti\ifmmode \acute{c}\else \'{c}\fi{}, Igor},
  journal = {Phys. Rev. Lett.},
  volume = {135},
  issue = {26},
  pages = {266703},
  numpages = {9},
  year = {2025},
  month = {Dec},
  publisher = {American Physical Society},
  doi = {10.1103/zn7r-k1xd},
  url = {https://link.aps.org/doi/10.1103/zn7r-k1xd}
}

@article{10.1063/1.1708514,
    author = {Brinkman, W. and Elliott, R. J.},
    title = {Space Group Theory for Spin Waves},
    journal = {Journal of Applied Physics},
    volume = {37},
    number = {3},
    pages = {1457-1459},
    year = {1966},
    month = {03},
    doi = {10.1063/1.1708514},
    url = {https://doi.org/10.1063/1.1708514}
}

@article{mkk5-b53m,
  title = {$d$-wave surface altermagnetism in centrosymmetric collinear antiferromagnets},
  author = {\ifmmode \mbox{\c{S}}\else \c{S}\fi{}a\ifmmode \mbox{\c{s}}\else \c{s}\fi{}\ifmmode \imath \else \i \fi{}o\ifmmode \breve{g}\else \u{g}\fi{}lu, Ersoy and Mertig, Ingrid and Lounis, Samir},
  journal = {Phys. Rev. B},
  volume = {114},
  issue = {2},
  pages = {L020406},
  numpages = {7},
  year = {2026},
  month = {Jul},
  publisher = {American Physical Society},
  doi = {10.1103/mkk5-b53m},
  url = {https://link.aps.org/doi/10.1103/mkk5-b53m}
}

@misc{leeb2026topologicallyprotectedsurfacealtermagnetism,
      title={Topologically Protected Surface Altermagnetism on Antiferromagnets}, 
      author={Valentin Leeb and Peru d'Ornellas and Fernando de Juan and Adolfo G. Grushin},
      year={2026},
      eprint={2602.10108},
      archivePrefix={arXiv},
      primaryClass={cond-mat.str-el},
      url={https://arxiv.org/abs/2602.10108}, 
}

@misc{lange2026emergentaltermagnetismsurfacesantiferromagnets,
      title={Emergent altermagnetism at surfaces of antiferromagnets: full symmetry classification and material identification}, 
      author={Colin Lange and Rodrigo Jaeschke-Ubiergo and Atasi Chakraborty and Xanthe H. Verbeek and Libor Šmejkal and Jairo Sinova and Alexander Mook},
      year={2026},
      eprint={2602.08773},
      archivePrefix={arXiv},
      primaryClass={cond-mat.mtrl-sci},
      url={https://arxiv.org/abs/2602.08773}, 
}

@article{hw4l-jknk,
  title = {Emergent surface altermagnetism},
  author = {Hu, Yuzhong and Zhou, Pan and Pan, Baoru and Liu, Songmin and Zhou, Binchang and Sun, Lizhong},
  journal = {Phys. Rev. Lett.},
  pages = {},
  year = {2026},
  month = {Aug},
  publisher = {American Physical Society},
  doi = {10.1103/hw4l-jknk},
  url = {https://link.aps.org/doi/10.1103/hw4l-jknk}
}

@ARTICLE{Litvin2008-ka,
  title    = {Tables of crystallographic properties of magnetic space groups},
  author   = {Litvin, D B},
  journal  = {Acta Crystallogr A},
  volume   =  {64},
  number   = {Pt 3},
  pages    = {419--424},
  month    =  {apr},
  year     =  {2008},
  address  = {United States}
}

@article{PhysRevX.14.031037,
  title = {Spin Space Groups: Full Classification and Applications},
  author = {Xiao, Zhenyu and Zhao, Jianzhou and Li, Yanqi and Shindou, Ryuichi and Song, Zhi-Da},
  journal = {Phys. Rev. X},
  volume = {14},
  issue = {3},
  pages = {031037},
  numpages = {33},
  year = {2024},
  month = {Aug},
  publisher = {American Physical Society},
  doi = {10.1103/PhysRevX.14.031037},
  url = {https://link.aps.org/doi/10.1103/PhysRevX.14.031037}
}

@article{PhysRevLett.132.176702,
  title = {Landau Theory of Altermagnetism},
  author = {McClarty, Paul A. and Rau, Jeffrey G.},
  journal = {Phys. Rev. Lett.},
  volume = {132},
  issue = {17},
  pages = {176702},
  numpages = {8},
  year = {2024},
  month = {Apr},
  publisher = {American Physical Society},
  doi = {10.1103/PhysRevLett.132.176702},
  url = {https://link.aps.org/doi/10.1103/PhysRevLett.132.176702}
}

@article{q44z-ynbr,
  title = {Collinear altermagnets and their Landau theories},
  author = {Schiff, Hana and McClarty, Paul and Rau, Jeffrey G. and Romh\'anyi, Judit},
  journal = {Phys. Rev. Res.},
  volume = {7},
  issue = {3},
  pages = {033301},
  numpages = {31},
  year = {2025},
  month = {Sep},
  publisher = {American Physical Society},
  doi = {10.1103/q44z-ynbr},
  url = {https://link.aps.org/doi/10.1103/q44z-ynbr}
}

@misc{sufyan2026,
      title={High-Throughput Quantification of Altermagnetic Band Splitting}, 
      author={Ali Sufyan and Brahim Marfoua and J. Andreas Larsson and Erik van Loon and Rickard Armiento},
      year={2026},
      eprint={2509.14729},
      archivePrefix={arXiv},
      primaryClass={cond-mat.mtrl-sci},
      url={https://arxiv.org/abs/2509.14729}, 
}

@article{LITVIN1974538,
title = {Spin groups},
journal = {Physica},
volume = {76},
number = {3},
pages = {538-554},
year = {1974},
issn = {0031-8914},
doi = {https://doi.org/10.1016/0031-8914(74)90157-8},
url = {https://www.sciencedirect.com/science/article/pii/0031891474901578},
author = {D.B. Litvin and W. Opechowski}
}

@article{Etxebarria:cam5007,
author = {Etxebarria, Jesus and Perez-Mato, J. Manuel and Tasci, Emre S. and Elcoro, Luis},
title = {Crystal tensor properties of magnetic materials with and without spin--orbit coupling. Application of spin point groups as approximate symmetries},
journal = {Acta Crystallographica Section A},
year = {2025},
volume = {81},
number = {4},
pages = {317-338},
month = {Jul},
doi = {10.1107/S2053273325004127},
url = {https://doi.org/10.1107/S2053273325004127},
}

@software{Cole_Python_Tight_Binding_2025,
author = {Cole, Trey and Coh, Sinisa and Vanderbilt, David},
doi = {10.5281/zenodo.12721315},
license = {GPL-3.0-or-later},
month = nov,
title = {{Python Tight Binding (PythTB)}},
url = {https://zenodo.org/records/12721315},
version = {2.0.0},
year = {2025}
}

@article{PhysRevB.47.558,
  title = {Ab initio molecular dynamics for liquid metals},
  author = {Kresse, G. and Hafner, J.},
  journal = {Phys. Rev. B},
  volume = {47},
  issue = {1},
  pages = {558--561},
  numpages = {0},
  year = {1993},
  month = {Jan},
  publisher = {American Physical Society},
  doi = {10.1103/PhysRevB.47.558},
  url = {https://link.aps.org/doi/10.1103/PhysRevB.47.558}
}

@article{PhysRevB.54.11169,
  title = {Efficient iterative schemes for ab initio total-energy calculations using a plane-wave basis set},
  author = {Kresse, G. and Furthm\"uller, J.},
  journal = {Phys. Rev. B},
  volume = {54},
  issue = {16},
  pages = {11169--11186},
  numpages = {0},
  year = {1996},
  month = {Oct},
  publisher = {American Physical Society},
  doi = {10.1103/PhysRevB.54.11169},
  url = {https://link.aps.org/doi/10.1103/PhysRevB.54.11169}
}

@article{PhysRevB.59.1758,
  title = {From ultrasoft pseudopotentials to the projector augmented-wave method},
  author = {Kresse, G. and Joubert, D.},
  journal = {Phys. Rev. B},
  volume = {59},
  issue = {3},
  pages = {1758--1775},
  numpages = {0},
  year = {1999},
  month = {Jan},
  publisher = {American Physical Society},
  doi = {10.1103/PhysRevB.59.1758},
  url = {https://link.aps.org/doi/10.1103/PhysRevB.59.1758}
}

@article{PhysRevB.87.235109,
  title = {Symmetry-adapted Wannier functions in the maximal localization procedure},
  author = {Sakuma, R.},
  journal = {Phys. Rev. B},
  volume = {87},
  issue = {23},
  pages = {235109},
  numpages = {8},
  year = {2013},
  month = {Jun},
  publisher = {American Physical Society},
  doi = {10.1103/PhysRevB.87.235109},
  url = {https://link.aps.org/doi/10.1103/PhysRevB.87.235109}
}

@article{PhysRevB.56.12847,
  title = {Maximally localized generalized Wannier functions for composite energy bands},
  author = {Marzari, Nicola and Vanderbilt, David},
  journal = {Phys. Rev. B},
  volume = {56},
  issue = {20},
  pages = {12847--12865},
  numpages = {0},
  year = {1997},
  month = {Nov},
  publisher = {American Physical Society},
  doi = {10.1103/PhysRevB.56.12847},
  url = {https://link.aps.org/doi/10.1103/PhysRevB.56.12847}
}

@article{PhysRevLett.134.196907,
  title = {Optical Signatures of Spin Symmetries in Unconventional Magnets},
  author = {Sivianes, Javier and Santos, Flaviano Jos\'e dos and Iba\~nez-Azpiroz, Julen},
  journal = {Phys. Rev. Lett.},
  volume = {134},
  issue = {19},
  pages = {196907},
  numpages = {7},
  year = {2025},
  month = {May},
  publisher = {American Physical Society},
  doi = {10.1103/PhysRevLett.134.196907},
  url = {https://link.aps.org/doi/10.1103/PhysRevLett.134.196907}
}

@misc{weber2026opticalexcitationspinpolarization,
      title={All optical excitation of spin polarization in d-wave altermagnets}, 
      author={Marius Weber and Stephan Wust and Luca Haag and Paul Herrgen and Akashdeep Akashdeep and Kai Leckron and Christin Schmitt and Rafael Ramos and Takashi Kikkawa and Eiji Saitoh and Mathias Kläui and Libor Šmejkal and Jairo Sinova and Martin Aeschlimann and Gerhard Jakob and Benjamin Stadtmüller and Hans Christian Schneider},
      year={2026},
      eprint={2408.05187},
      archivePrefix={arXiv},
      primaryClass={cond-mat.mtrl-sci},
      url={https://arxiv.org/abs/2408.05187}, 
}

@misc{vila2024orbitalspinlockingopticalsignatures,
      title={Orbital-spin Locking and its Optical Signatures in Altermagnets}, 
      author={Marc Vila and Veronika Sunko and Joel E. Moore},
      year={2024},
      eprint={2410.23513},
      archivePrefix={arXiv},
      primaryClass={cond-mat.mtrl-sci},
      url={https://arxiv.org/abs/2410.23513}, 
}

@Article{Farhang2026,
author={Farhang, Camron
and Lu, Weihang
and Du, Kai
and Gao, Yunpeng
and Yang, Junjie
and Cheong, Sang-Wook
and Xia, Jing},
title={Topological magneto-optical Kerr effect without spin-orbit coupling in spin-compensated antiferromagnet},
journal={Nature Communications},
year={2026},
month={Mar},
day={03},
volume={17},
number={1},
pages={3386},
issn={2041-1723},
doi={10.1038/s41467-026-70238-0},
url={https://doi.org/10.1038/s41467-026-70238-0}
}

@article{463y-q7lt,
  title = {Quantization of Spin Circular Photogalvanic Effect in Altermagnetic Weyl Semimetals},
  author = {Yoshida, Hiroki and Priessnitz, Jan and \ifmmode \check{S}\else \v{S}\fi{}mejkal, Libor and Murakami, Shuichi},
  journal = {Phys. Rev. Lett.},
  volume = {136},
  issue = {9},
  pages = {096701},
  numpages = {9},
  year = {2026},
  month = {Mar},
  publisher = {American Physical Society},
  doi = {10.1103/463y-q7lt},
  url = {https://link.aps.org/doi/10.1103/463y-q7lt}
}

@Article{Feng2020,
author={Feng, Wanxiang
and Hanke, Jan-Philipp
and Zhou, Xiaodong
and Guo, Guang-Yu
and Bl{\"u}gel, Stefan
and Mokrousov, Yuriy
and Yao, Yugui},
title={Topological magneto-optical effects and their quantization in noncoplanar antiferromagnets},
journal={Nature Communications},
year={2020},
month={Jan},
day={08},
volume={11},
number={1},
pages={118},
issn={2041-1723},
doi={10.1038/s41467-019-13968-8},
url={https://doi.org/10.1038/s41467-019-13968-8}
}

@article{10.1021/acs.nanolett.5c03647,
    author = {Sun, Jiuyu and Du, Yongping and Kan, Erjun},
    title = {Symmetry-Breaking
Magneto-Optical Effects in Altermagnets},
    journal = {Nano Letters},
    volume = {25},
    number = {41},
    pages = {14960-14966},
    year = {2025},
    month = {10},
    issn = {1530-6984},
    doi = {10.1021/acs.nanolett.5c03647},
    url = {https://doi.org/10.1021/acs.nanolett.5c03647},
    eprint = {https://pubs.acs.org/nalefd/article-pdf/25/41/14960/42225892/nl5c03647.pdf},
}

@article{q8ym-l2zt,
  title = {Experimental Evidence of N\'eel-Order-Driven Magneto-optical Kerr Effect in an Altermagnetic Insulator},
  author = {Pan, Haolin and Xiao, Rui-Chun and Han, Jiahao and Zhu, Hongxing and Li, Junxue and Niu, Qian and Gao, Yang and Hou, Dazhi},
  journal = {Phys. Rev. Lett.},
  volume = {136},
  issue = {3},
  pages = {036701},
  numpages = {7},
  year = {2026},
  month = {Jan},
  publisher = {American Physical Society},
  doi = {10.1103/q8ym-l2zt},
  url = {https://link.aps.org/doi/10.1103/q8ym-l2zt}
}

@article{PhysRevLett.132.176701,
  title = {X-Ray Magnetic Circular Dichroism in Altermagnetic $\ensuremath{\alpha}$-MnTe},
  author = {Hariki, A. and Dal Din, A. and Amin, O. J. and Yamaguchi, T. and Badura, A. and Kriegner, D. and Edmonds, K. W. and Campion, R. P. and Wadley, P. and Backes, D. and Veiga, L. S. I. and Dhesi, S. S. and Springholz, G. and \ifmmode \check{S}\else \v{S}\fi{}mejkal, L. and V\'yborn\'y, K. and Jungwirth, T. and Kune\ifmmode \check{s}\else \v{s}\fi{}, J.},
  journal = {Phys. Rev. Lett.},
  volume = {132},
  issue = {17},
  pages = {176701},
  numpages = {7},
  year = {2024},
  month = {Apr},
  publisher = {American Physical Society},
  doi = {10.1103/PhysRevLett.132.176701},
  url = {https://link.aps.org/doi/10.1103/PhysRevLett.132.176701}
}

@article{KIMEL2024172039,
title = {Optical read-out and control of antiferromagnetic Néel vector in altermagnets and beyond},
journal = {Journal of Magnetism and Magnetic Materials},
volume = {598},
pages = {172039},
year = {2024},
issn = {0304-8853},
doi = {https://doi.org/10.1016/j.jmmm.2024.172039},
url = {https://www.sciencedirect.com/science/article/pii/S0304885324003305},
author = {A.V. Kimel and Th. Rasing and B.A. Ivanov}
}

@Article{Sunko2026,
author={Sunko, V.
and Orenstein, J.},
title={Linear magneto-birefringence as a probe of altermagnetism},
journal={npj Quantum Materials},
year={2026},
month={May},
day={30},
issn={2397-4648},
doi={10.1038/s41535-026-00901-8},
url={https://doi.org/10.1038/s41535-026-00901-8}
}

@article{PhysRevB.110.024425,
  title = {Tunable band topology and optical conductivity in altermagnets},
  author = {Rao, Peng and Mook, Alexander and Knolle, Johannes},
  journal = {Phys. Rev. B},
  volume = {110},
  issue = {2},
  pages = {024425},
  numpages = {12},
  year = {2024},
  month = {Jul},
  publisher = {American Physical Society},
  doi = {10.1103/PhysRevB.110.024425},
  url = {https://link.aps.org/doi/10.1103/PhysRevB.110.024425}
}

@Article{Weber2025,
author={Weber, Marius
and Leckron, Kai
and Haag, Luca Felipe
and Jaeschke-Ubiergo, Rodrigo
and {\v{S}}mejkal, Libor
and Sinova, Jairo
and Schneider, Hans Christian},
title={Ultrafast electron dynamics in a planar d-wave altermagnet},
journal={Newton},
year={2025},
month={Dec},
day={01},
publisher={Elsevier},
volume={1},
number={10},
issn={2950-6360},
doi={10.1016/j.newton.2025.100266},
url={https://doi.org/10.1016/j.newton.2025.100266}
}

@article{pizzi_wannier90_2019,
                title = {Wannier90 as a community code: new features and applications},
        issn = {0953-8984},
        url = {http://iopscience.iop.org/10.1088/1361-648X/ab51ff},
        doi = {10.1088/1361-648X/ab51ff},
        journal = {Journal of Physics: Condensed Matter},
        shortjournal = {J. Phys.: Condens. Matter},
        author = {Pizzi, Giovanni and Vitale, Valerio and Arita, Ryotaro and Bluegel, Stefan and Freimuth, Frank and Géranton, Guillaume and Gibertini, Marco and Gresch, Dominik and Johnson, Charles and Koretsune, Takashi and Ibanez, Julen and Lee, Hyungjun and Lihm, Jae-Mo and Marchand, Daniel and Marrazzo, Antimo and Mokrousov, Yuriy and Mustafa, Jamal Ibrahim and Nohara, Yoshiro and Nomura, Yusuke and Paulatto, Lorenzo and Ponce, Samuel and Ponweiser, Thomas and Qiao, Junfeng and Thöle, Florian and Tsirkin, Stepan S. and Wierzbowska, Malgorzata and Marzari, Nicola and Vanderbilt, David and Souza, Ivo and Mostofi, Arash A. and Yates, Jonathan R.},
        year = {2019}
        }

@Article{Tsirkin2021,
author={Tsirkin, Stepan S.},
title={High performance {W}annier interpolation of Berry curvature and related quantities with {WannierBerri} code},
journal={npj Computational Materials},
year={2021},
month={Feb},
day={19},
volume={7},
number={1},
pages={33},
issn={2057-3960},
doi={10.1038/s41524-021-00498-5},
url={https://doi.org/10.1038/s41524-021-00498-5}
}

@article{Sinova2015,
  title = {Spin {H}all effects},
  author = {Sinova, Jairo and Valenzuela, Sergio O. and Wunderlich, J. and Back, C. H. and Jungwirth, T.},
  journal = {Rev. Mod. Phys.},
  volume = {87},
  issue = {4},
  pages = {1213--1260},
  numpages = {47},
  year = {2015},
  month = {Oct},
  publisher = {American Physical Society},
  doi = {10.1103/RevModPhys.87.1213},
  url = {https://link.aps.org/doi/10.1103/RevModPhys.87.1213}
}

@article{Perdew1996,
  title = {Generalized Gradient Approximation Made Simple},
  author = {Perdew, John P. and Burke, Kieron and Ernzerhof, Matthias},
  journal = {Phys. Rev. Lett.},
  volume = {77},
  issue = {18},
  pages = {3865--3868},
  numpages = {0},
  year = {1996},
  month = {Oct},
  publisher = {American Physical Society},
  doi = {10.1103/PhysRevLett.77.3865},
  url = {https://link.aps.org/doi/10.1103/PhysRevLett.77.3865}
}

@article{Ryoo2019,
  title = {Computation of intrinsic spin {H}all conductivities from first principles using maximally localized {W}Siannier functions},
  author = {Ryoo, Ji Hoon and Park, Cheol-Hwan and Souza, Ivo},
  journal = {Phys. Rev. B},
  volume = {99},
  issue = {23},
  pages = {235113},
  numpages = {11},
  year = {2019},
  month = {Jun},
  publisher = {American Physical Society},
  doi = {10.1103/PhysRevB.99.235113},
  url = {https://link.aps.org/doi/10.1103/PhysRevB.99.235113}
}

@book{Landau1984,
  author    = {Landau, L. D. and Lifshitz, E. M. and Pitaevskii, L. P.},
  title     = {Electrodynamics of Continuous Media},
  series    = {Course of Theoretical Physics},
  volume    = {8},
  edition   = {2},
  publisher = {Pergamon Press},
  address   = {Oxford},
  year      = {1984},
  isbn      = {9780080302751}
}

@article{cheong2024altermagnetism,
  title   = {Altermagnetism with non-collinear spins},
  author  = {Cheong, Sang-Wook and Huang, Fei-Ting},
  journal = {npj Quantum Materials},
  volume  = {9},
  number  = {1},
  pages   = {13},
  year    = {2024},
  doi     = {10.1038/s41535-024-00626-6}
}

@article{angle,
    author = {Sturm, C. and Furthmüller, J. and Bechstedt, F. and Schmidt-Grund, R. and Grundmann, M.},
    title = {Dielectric tensor of monoclinic Ga2O3 single crystals in the spectral range 0.5–8.5 eV},
    journal = {APL Materials},
    volume = {3},
    number = {10},
    pages = {106106},
    year = {2015},
    month = {10},
    issn = {2166-532X},
    doi = {10.1063/1.4934705},
    url = {https://doi.org/10.1063/1.4934705}
}

@article{ratio,
author = {Wang, Tianyu and Zhao, Kai and Wang, Pan and Shen, Wanfu and Gao, Haikuo and Qin, Zhengsheng and Wang, Yongshuai and Li, Chunlei and Deng, Huixiong and Hu, Chunguang and Jiang, Lang and Dong, Huanli and Wei, Zhongming and Li, Liqiang and Hu, Wenping},
title = {Intrinsic Linear Dichroism of Organic Single Crystals toward High-Performance Polarization-Sensitive Photodetectors},
journal = {Advanced Materials},
volume = {34},
number = {22},
pages = {2105665},
doi = {https://doi.org/10.1002/adma.202105665},
url = {https://advanced.onlinelibrary.wiley.com/doi/abs/10.1002/adma.202105665},
year = {2022}
}

@book{10.1093/oso/9780198505921.001.0001,
    author = {Blundell, Stephen},
    title = {Magnetism in Condensed Matter},
    publisher = {Oxford University Press},
    year = {2001},
    month = {10},
    isbn = {9780198505921},
    doi = {10.1093/oso/9780198505921.001.0001},
    url = {https://doi.org/10.1093/oso/9780198505921.001.0001},
}

@inbook{Lifshitz_2024,
   title={Magnetic point groups and space groups},
   ISBN={9780323914086},
   url={http://dx.doi.org/10.1016/B978-0-323-90800-9.00082-2},
   DOI={10.1016/b978-0-323-90800-9.00082-2},
   booktitle={Encyclopedia of Condensed Matter Physics},
   publisher={Elsevier},
   author={Lifshitz, Ron},
   year={2024},
   pages={1–10} }

@book{chakraborty2023encyclopedia,
  editor    = {Tapash Chakraborty},
  title     = {Encyclopedia of Condensed Matter Physics},
  edition   = {2},
  publisher = {Elsevier},
  address   = {Amsterdam, Netherlands},
  year      = {2023},
  month     = oct,
  isbn      = {978-0-323-90800-9},
  note      = {Hardback ISBN: 9780323908009, eBook ISBN: 9780323914086},
  url       = {https://shop.elsevier.com/books/encyclopedia-of-condensed-matter-physics/chakraborty/978-0-323-90800-9}
}

@article{PhysRevB.104.235403,
  title = {Anisotropic magneto-optical absorption and linear dichroism in two-dimensional semi-Dirac electron systems},
  author = {Zhou, Xiaoying and Chen, Wang and Zhu, Xianzhe},
  journal = {Phys. Rev. B},
  volume = {104},
  issue = {23},
  pages = {235403},
  numpages = {9},
  year = {2021},
  month = {Dec},
  publisher = {American Physical Society},
  doi = {10.1103/PhysRevB.104.235403},
  url = {https://link.aps.org/doi/10.1103/PhysRevB.104.235403}
}

@article{
doi:10.1126/science.1248552,
author = {E. R. Schemm  and W. J. Gannon  and C. M. Wishne  and W. P. Halperin  and A. Kapitulnik },
title = {Observation of broken time-reversal symmetry in the heavy-fermion superconductor UPt<sub>3</sub>},
journal = {Science},
volume = {345},
number = {6193},
pages = {190-193},
year = {2014},
doi = {10.1126/science.1248552},
URL = {https://www.science.org/doi/abs/10.1126/science.1248552},
eprint = {https://www.science.org/doi/pdf/10.1126/science.1248552}}

@article{PhysRevB.57.1505,
  title = {Electron-energy-loss spectra and the structural stability of nickel oxide:  An LSDA+U study},
  author = {Dudarev, S. L. and Botton, G. A. and Savrasov, S. Y. and Humphreys, C. J. and Sutton, A. P.},
  journal = {Phys. Rev. B},
  volume = {57},
  issue = {3},
  pages = {1505--1509},
  numpages = {0},
  year = {1998},
  month = {Jan},
  publisher = {American Physical Society},
  doi = {10.1103/PhysRevB.57.1505},
  url = {https://link.aps.org/doi/10.1103/PhysRevB.57.1505}
}

@article{mmdm-hrj4,
  title = {High-throughput quantification of altermagnetic band splitting},
  author = {Sufyan, Ali and Marfoua, Brahim and Larsson, J. Andreas and van Loon, Erik and Armiento, Rickard},
  journal = {Phys. Rev. Mater.},
  volume = {10},
  issue = {4},
  pages = {044407},
  numpages = {11},
  year = {2026},
  month = {Apr},
  publisher = {American Physical Society},
  doi = {10.1103/mmdm-hrj4},
  url = {https://link.aps.org/doi/10.1103/mmdm-hrj4}
}

@article{Gallego:ks5532,
author = "Gallego, Samuel V. and Perez-Mato, J. Manuel and Elcoro, Luis and Tasci, Emre S. and Hanson, Robert M. and Momma, Koichi and Aroyo, Mois I. and Madariaga, Gotzon",
title = "{{\it MAGNDATA}: towards a database of magnetic structures. I.The commensurate case}",
journal = "Journal of Applied Crystallography",
year = "2016",
volume = "49",
number = "5",
pages = "1750--1776",
month = "Oct",
doi = {10.1107/S1600576716012863},
url = {https://doi.org/10.1107/S1600576716012863},
}

@article{Gallego:ks5530,
author = "Gallego, Samuel V. and Perez-Mato, J. Manuel and Elcoro, Luis and Tasci, Emre S. and Hanson, Robert M. and Aroyo, Mois I. and Madariaga, Gotzon",
title = "{{\it MAGNDATA}: towards a database of magnetic structures. II. The incommensurate case}",
journal = "Journal of Applied Crystallography",
year = "2016",
volume = "49",
number = "6",
pages = "1941--1956",
month = "Dec",
doi = {10.1107/S1600576716015491},
url = {https://doi.org/10.1107/S1600576716015491},
}

@article{10.1021/jacs.7b06314,
    author = {Wang, Xiaoting and Li, Yongtao and Huang, Le and Jiang, Xiang-Wei and Jiang, Lang and Dong, Huanli and Wei, Zhongming and Li, Jingbo and Hu, Wenping},
    title = {Short-Wave
Near-Infrared Linear Dichroism of Two-Dimensional
Germanium Selenide},
    journal = {Journal of the American Chemical Society},
    volume = {139},
    number = {42},
    pages = {14976-14982},
    year = {2017},
    month = {09},
    issn = {0002-7863},
    doi = {10.1021/jacs.7b06314},
    url = {https://doi.org/10.1021/jacs.7b06314},
    eprint = {https://pubs.acs.org/jacsat/article-pdf/139/42/14976/15464114/ja7b06314.pdf},
}

@article{Zhong:21,
author = {Jiahong Zhong and Cheng Zeng and Juan Yu and Lingkai Cao and Junnan Ding and Zongwen Liu and Yanping Liu},
journal = {Opt. Express},
number = {3},
pages = {3567--3574},
publisher = {Optica Publishing Group},
title = {Direct observation of enhanced performance in suspended ReS2 photodetectors},
volume = {29},
month = {Feb},
year = {2021},
url = {https://opg.optica.org/oe/abstract.cfm?URI=oe-29-3-3567},
doi = {10.1364/OE.415060},
}

@article{doi:10.1143/JPSJ.12.570,
author = {Kubo ,Ryogo},
title = {Statistical-Mechanical Theory of Irreversible Processes. I. General Theory and Simple Applications to Magnetic and Conduction Problems},
journal = {Journal of the Physical Society of Japan},
volume = {12},
number = {6},
pages = {570-586},
year = {1957},
doi = {10.1143/JPSJ.12.570},

URL = { 
    
        https://doi.org/10.1143/JPSJ.12.570
    
    

},
eprint = { 
    
        https://doi.org/10.1143/JPSJ.12.570
    
    

}
}

@article{Greenwood_1958,
doi = {10.1088/0370-1328/71/4/306},
url = {https://doi.org/10.1088/0370-1328/71/4/306},
year = {1958},
month = {apr},
publisher = {},
volume = {71},
number = {4},
pages = {585},
author = {D A Greenwood},
title = {The Boltzmann Equation in the Theory of Electrical Conduction in Metals},
journal = {Proceedings of the Physical Society}
}
\end{document}